\documentclass[manuscript]{aa} 

\usepackage[varg]{txfonts}
\usepackage{graphicx}
\usepackage{amssymb, amsmath} 
\usepackage{color}
\usepackage[usenames,dvipsnames,table]{xcolor}
\bibpunct{(}{)}{;}{a}{}{,} 

\def \deg{^\circ}

\def \ferg {\mbox{erg cm$^{-2}$ s$^{-1}$}}
\def \hcm {\hbox {\ifmmode $ atom cm$^{-2}\else atom cm$^{-2}$\fi}}
\def \chisq {$\chi ^{2}$}
\def \Halpha{H-$\alpha$}

\def \msun {$M_{\odot}$}

\def\chan{{\em Chandra}}

\def\j11{\mbox{\object{IGR~J11014-6103}}}
\def\lhn{Lighthouse nebula} 
\def\SNR{\object{SNR MSH 11-61A}}

\def\morla{\object{PSR J0357+3205}}

\def \apjl {ApJL}
\def \apjs {ApJS}
\def \aap {A\&A}

\begin{document}

\title{A closer view of the \j11\ outflows} \titlerunning{\j11\ outflows}

\author {L. Pavan\inst{\ref{inst1}} 
  \and  G. P{\"u}hlhofer\inst{\ref{inst2}} 
  \and P. Bordas\inst{\ref{inst3}}
  \and M. Audard\inst{\ref{inst1}} 
  \and M. Balbo\inst{\ref{inst1}}
  \and E. Bozzo\inst{\ref{inst1}} 
  \and D. Eckert\inst{\ref{inst1}}
  \and C. Ferrigno\inst{\ref{inst1}} 
  \and  M. D. Filipovi\'c\inst{\ref{inst4}} 
  \and M. Verdugo\inst{\ref{inst5}}
  \and R. Walter\inst{\ref{inst1}} }

\institute{  Universit\'e de Gen\`eve, Departement d'Astronomie -- ISDC,
  chemin d'Ecogia, 16, 1290 Versoix, Switzerland\\
  \email{Lucia.Pavan@unige.ch}
\label{inst1} 
\and Institut f{\"u}r
  Astronomie und Astrophysik, Universit{\"a}t T{\"u}bingen, Sand 1,
  D-72076, T{\"u}bingen, Germany
\label{inst2} 
\and Max-Planck-Institut
  f{\"u}r Kernphysik Saupfercheckweg 1, D-69117, Heidelberg,
  Germany
\label{inst3} 
\and Western Sydney University, Locked Bag 1797, Penrith South DC, NSW
1797, Australia
\label{inst4}
\and Department for Astrophysics,
  University of Vienna T\"urkenschanzstr. 17, 1180 Vienna,
  Austria
\label{inst5} 
}

\authorrunning{L.Pavan et al.}

\date{Received / Accepted }

\abstract{
  { The complex X-ray system \j11 (a.k.a. the \lhn)
  is composed of a bow-shock pulsar wind nebula (PWN) as well as large-scale
  jet-like features, all launched by \j11\ which is moving supersonically in
  the interstellar medium. Previous observations suggested that the jet features stem
  from a ballistic jet of relativistic particles.}
{In order to confirm the nature of the jet and the marginally detected counter-jet,
 we obtained a new deep 250~ks \chan\ observation of the \lhn. 
  } 
{We performed detailed spatial and spectral analysis of all X-ray
  components of the system.
}
{The X-ray PWN is now better resolved and shows a clear bi-modal
  morphology.  The overall helical pattern of the main jet is
  confirmed. However, there are large deviations from a simple helical
  model at small and large scales.  Significant extended emission is
  now detected, encompassing the main jet all along its length.  The
  brightness dip of the main jet at $\sim 50\arcsec$ distance from the
  pulsar is confirmed, the extended emission however prevents conclusions
  about the coherence at this position of the jet.  The counter-jet is
  now detected at high statistical significance. In addition, we found
  two small-scale ``arcs'' departing from the pulsar towards the jets.
  We also looked for possible bow-shock emission in front of
  the pulsar, with a short VLT/FORS2 \Halpha\ observation. 
  No clear emission is found, most likely because of contamination from a
  diffuse surrounding nebulosity.}
{The results of our X-ray analysis show that current expectations from a
  ballistic nature of the jets can explain satisfactorily some of the
  observational evidences but cannot fully reproduce the
  observations. The alternative scenario (diffusion of particles along
  preexisting magnetic field lines in the surrounding medium) however
  also continues to suffer from conflicts with the observations.}  }


\maketitle

\defcitealias{pavan11}{Paper I} \defcitealias{pavan13}{Paper II}


\section{Introduction}
\label{sec:introduction}

Neutron stars are one of the possible end points of the stellar
evolution, and are produced in most cases through the degenerate collapse of the
core of a massive star (once the internal part of the star has reached
an unstable condition). This core-collapse supernova explosion can
imprint a strong natal kick to the newly born neutron star, which can
consequently move with high velocity into the ambient medium \citep[see
e.g. the review by][] {Janka-2012}.  Detailed analyses of the galactic
pulsar population found a mean velocity of $\sim$400~km/s
\citep{hobbs2005}. The highest velocity directly
measured is that of PSR~B1508+55 with
$v_{\textrm{PSR}} = 1083^{+103}_{-90}$~km/s \citep{chatterjee:2005fk}.

When the spatial velocity of a pulsar is sufficiently large, it will
escape its progenitor's supernova remnant (SNR) while the SNR is still
young and active e.g. in X-rays.  The pulsar's motion in the interstellar
medium ISM is then typically supersonic.  In this case, a bow-shock is
generated in front of the pulsar, and the swept-up matter prevents the
wind expelled by the pulsar to propagate in the direction of motion.
This wind is then collimated backwards, forming a nebula confined in a
conical shape \citep[whose aperture angle depends, among other
parameters, on the pulsar linear velocity; for a review see,
e.g.][]{gaensler2005}.

MSH~11-61A (a.k.a. G290.1-0.8) is a mixed morphology SNR detected from
radio to soft X-rays (up to $\sim$3 keV), formed by the core collapse
of a massive progenitor star \citep[mass $\gtrsim 25$\msun;
][]{filipovic2005,reynoso2006, Garcia:2012fk, Kamitsukasa:2014aa,
  auchettl2015b}.  Following these authors, the distance to the SNR is
in the range 6 -11 kpc, with the latest values
converging towards $7 \pm 1$~kpc.
This distance of 7~kpc is adopted throughout the paper\footnote{for
  reference, at the distance of 7~kpc an angular scale of 1 arcmin
  corresponds to $\sim$ 2 pc.}.
Close to this SNR is the INTEGRAL source \j11\ -- an X-ray system
powered by the pulsar PSR~J1101-6101 and comprising several extended
structures (\citealp{pavan11}; \citealp{tomsick:2012eu};
\citealp{pavan13}, hereafter Paper~II; \citealp{Halpern2014}).
Previous \chan\ observations that were aimed at the INTEGRAL source identified
these structures as three outflows produced by PSR J1101-6101: an
X-ray and radio PWN, shaped in a narrow cone elongated towards the
parent SNR, and an X-ray jet and counter-jet, both oriented nearly
perpendicular to the PWN axis (\citealp{tomsick:2012eu},
\citetalias{pavan13}).  The main jet extends for nearly 5\arcmin\ in
the sky, which corresponds to a length of $\sim 11$ pc projected on
the plane of the sky, at the 7~kpc distance, and showed a remarkable
helicoidal pattern \citepalias[see ][]{pavan13}. 
Already in this data set, indications for a spatial deviation from the
helical pattern of the main jet were present at $\sim 50\arcsec$
distance from the pulsar. At this position the surface brightness of
the jet is low (seemingly a ``gap''), but the brightness profile was
compatible with expectations due to Doppler-deboosting in the
jet-helix model. The spatial deviation was therefore not considered
significant at the time, also because the data were hampered by the
presence of CCD chip gaps, resulting in only 50\% effective exposure
in that region.  The spatial deviation was therefore not considered
significant at the time.  

The counterjet was detected at 3.7$\sigma$
in the \chan\ image and characterised by a flux of $\sim$ 5\% 
that of the main jet.
The conical shape of the PWN in \j11\ is ascribed to the
supersonic motion of PSR J1101-6101 in the ISM \citep{tomsick:2012eu}.
Although several examples of bow-shock pulsar wind nebulae (bsPWNe)
have now been detected in connection with their pulsars travelling at
supersonic velocities in the interstellar medium \citep[see
e.g.][]{mouse2004, mushroomPWN, kargalstev-pavlov-2008}, only one other
system is sharing a geometry similar to the one seen in the \lhn: the
Guitar nebula, powered by PSR~B2224+65 \citep{Cordes:1993fk,
  Chatterjee:2002lr, johnson-wang-10, Hui:2012fk}.  In this source the
pulsar is powering a bright and elongated X-ray jet as well, extending
over $\sim$1~pc in a direction almost perpendicular to pulsar direction
of motion.
In both cases, that of the Lighthouse nebula and that of the Guitar
nebula, the mechanism leading to the production of such peculiar jets
is still poorly understood (\citealp{bandiera2008},
\citealp{johnson-wang-10}, \citealp{pavan11}, \citealp{Hui:2012fk},
\citetalias{pavan13}).

To solve a series of questions that could not be conclusively
addressed with the previous data set, we obtained in 2014 a much
deeper \chan\ X-ray observation (250~ks), the analysis of which we describe
here in Sect.~\ref{sec:observations}.  In contrast to our first \chan\
observation that was aimed at imaging with the highest accuracy the
pulsar and the PWN, we centered this second observation on the main jet
to investigate its intriguing morphology.  The observation was split
into 5 smaller exposures (with different roll angles) because of
planning constraints.  We then chose to optimise the pointing
displacement to have the main jet (and the pulsar) entirely included in a
single (and same) CCD, in order to optimally investigate the ``gap''
region along the main jet, and the region connecting the pulsar and the
jet base.  This was achieved with a simZ offset of 6.0~mm and a Y
offset of 1.0\arcmin. The offsets were kept identical in all exposures
to ensure a good handling of the combined events.
We report here also on the analysis of an exploratory observation
performed with VLT to search for possible \Halpha\ emission close to
the pulsar (Sect.~\ref{sec:vlt}).  In Sect.~\ref{sec:discussion} we
discuss the results of our data analyses, and provide our conclusions
in Sect.~\ref{sec:conclusions}.

\section{\chan\ X-ray observations and data analysis}
\label{sec:observations}

The new \chan\ observation was split into 5 shorter exposures, as
detailed in Table~\ref{tab:observations}.
The data have been processed with \chan\ CIAO v.4.7, using the latest
available calibrations (CALDB v. 4.6.7).
All observations have been reprocessed with the CIAO tool
\texttt{chandra\_repro},
using \texttt{vfaint} mode background cleaning and subpixel
resolution (\texttt{edser} method), unless differently specified.
After cleaning, we verified that this filter did not affect
significantly the event counts of the emitting structures in
\j11. Only the pulsar was partially affected by this filter, with
$\sim$6\%
of good events being rejected.  We used therefore the vfaint
background cleaned evt2 files in the following. We also verified that
the observations were not affected by high flaring background.  The
total cleaned exposure time was 247~ks.

By computing and inspecting the PSF map of each observation, we 
verified that in all cases the characteristics of the instrument PSF
were virtually constant in the region including the pulsar, the jets,
and the PWN.

\begin{figure*}[!htb]
  \includegraphics[width=\textwidth]{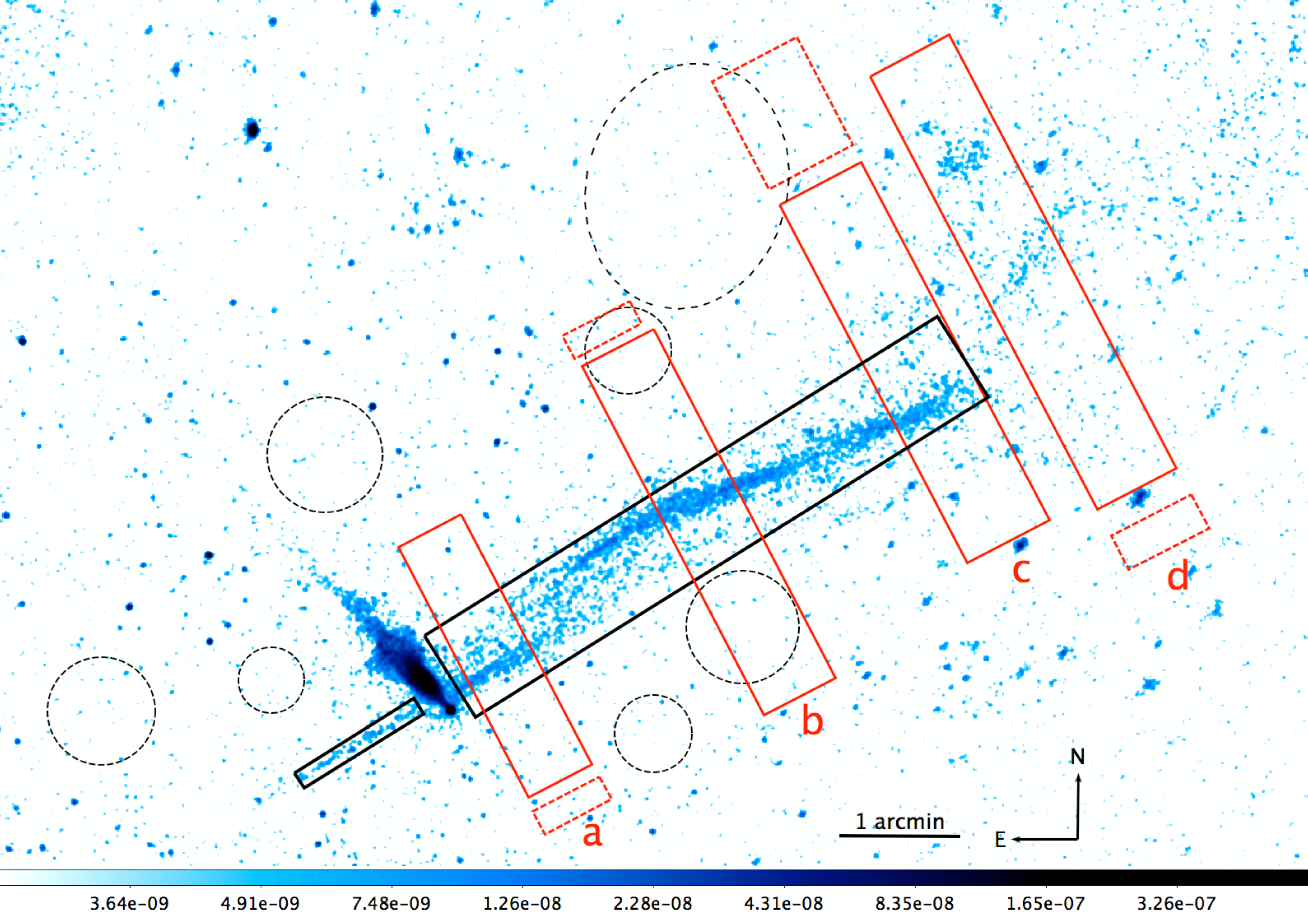}%
  \begin{picture}(0,0)
    \put(-530,220){\fbox{\includegraphics[height=5.5cm]{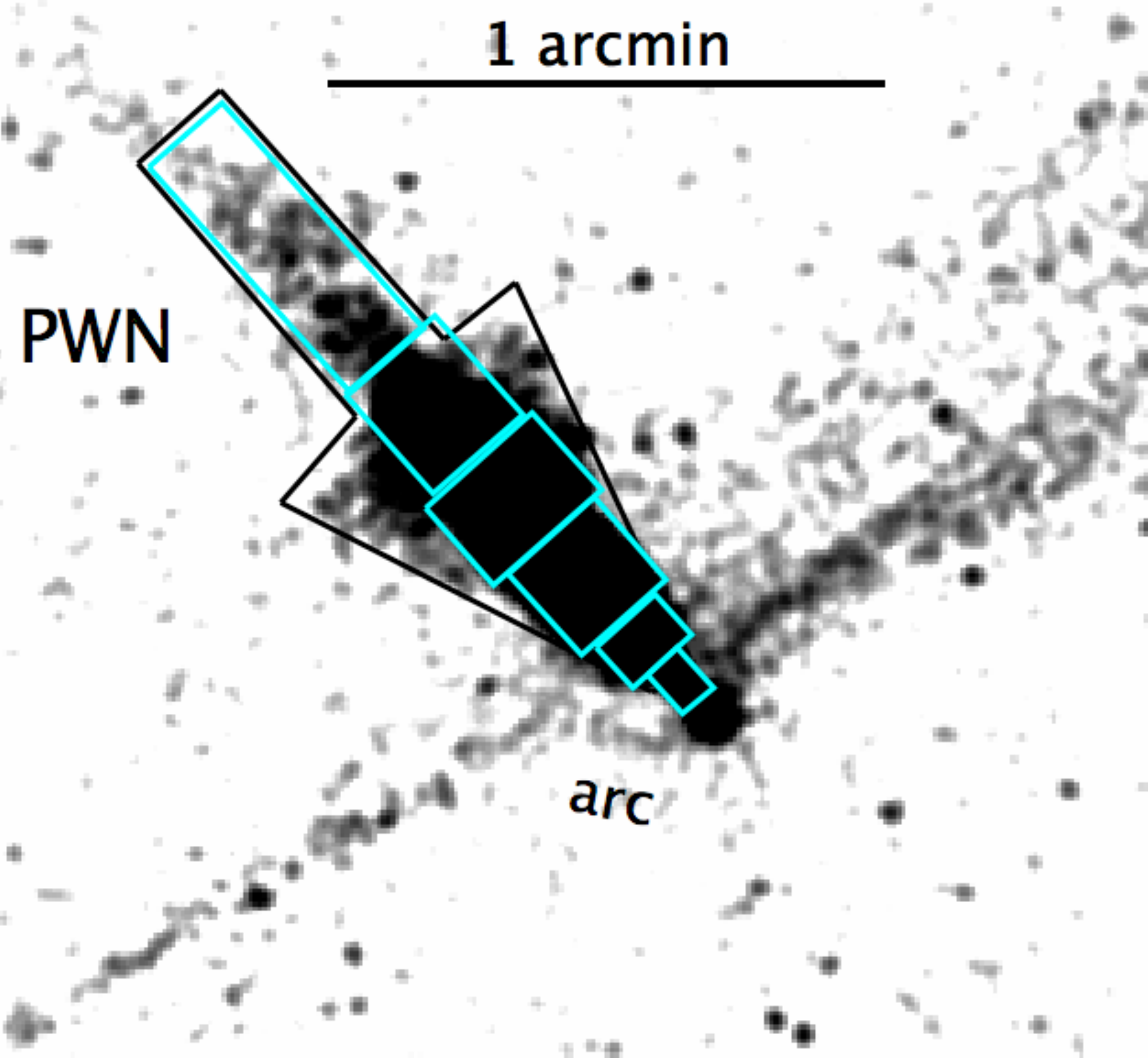}}}
  \end{picture}
  \caption{\chan\ 250~ks mosaic of the \lhn\
    (exposure corrected and smoothed with a gaussian kernel of $\sigma$=1.5 pixel). The
    color scale at the bottom of the plot is in units of
    photon/cm$^2$/s. 
    Solid red rectangles ``a''--``d'' (and the corresponding dashed background boxes)
    were used to extract brightness profiles perpendicularly to the
    main jet (Sect.~\ref{sec:briprof}).
    The regions used for spectral extraction are
    also shown: solid black regions are used for
    the extraction of the mean spectra of the main jet, the counter-jet and
    the PWN (Sect.~\ref{sec:mean-spectra}). For the point source PSR~J1101-6101 we used a circle of radius
    1.6\arcsec centered on the pulsar position (not displayed).
    Background regions are marked with dashed black circles. In the inset: the details of the PWN shape and the
    attaching points of the two jets to the pulsar are visible. Colored
    rectangular regions along the 
    PWN were used to extract spatially resolved spectra (Sect.~\ref{sec:spectra-in-space}).
  }
  \label{fig:j11}
\end{figure*}

\begin{table}[tb]
  \centering
    \caption{Summary of the \chan\ observations of \j11}
    \begin{tabular}{cccc}
      Obs ID &   Exposure &    Roll Angle &  Obs  Date\\ 
             & (ks)  &  (deg) & \\
      \hline\\
      \multicolumn{4}{l}{\it new observation}\\
      16007 & 116.0 &  189.2 & 2014-08-28\\
      16517 &  52.0 &  181.5 & 2014-09-05\\
      16518 &  10.0 &  155.2 & 2014-09-29\\
      17422 &  49.4 &  155.2 & 2014-10-01\\
      17421 &  19.8 &  155.2 & 2014-10-02\\
      \vspace{1pt}\\
      \multicolumn{4}{l}{\it previous obs}\\
      13787 &  49.4 &  142.3 & 2012-10-11\\
    \end{tabular}
  \label{tab:observations}
\end{table}

\subsection{Imaging}
\label{sec:imaging}

The new observations were combined in an exposure
corrected mosaic with the \texttt{ciao} tool \texttt{merge\_obs}, in
the energy range 0.5-7~keV (see Fig.~\ref{fig:j11}).  The main spatial
features seen in this mosaic are:

{\bf PWN:}
The PWN has a sharp conical shape, but it also shows a more
extended and collimated component. The overall structure resembles the
shape of an arrow, with the ``shaft'' of the arrow extending over
$\sim$1.7\arcmin\ in a compact cylindrical shape, and the ``head'' 
 extending on a wider and slightly dimmer cone, with an aperture of
 $\sim30\deg$ up to
$\sim 0.7$\arcmin\ from the pulsar (see inset in Fig.~\ref{fig:j11}). 
Comparing the new
image to our previous 50~ks observation, we note that the same
structure was present also in that observation, but less evident
because of the lower statistics.

{\bf counter-jet:}
The counter-jet is now very significantly detected and it extends
remarkably linearly for 1.5\arcmin, in SE direction. 
 It also appears to be roughly
aligned with the first 50\arcsec\ of the main jet.

{\bf Main jet:} Overall the main jet still shows the same pattern
found in our previous shorter observation \citepalias{pavan13},
including a spatial ``break'' at about 50-90\arcsec\ from the pulsar.
Whereas in the previous observation this region was coincident with a
factor two lower exposure due to the presence of ACIS chip gaps, the
new data have equal exposure along the jet, without chip gaps or other
significant dead chip columns.  We can thus confirm that the loss of a
coherent line structure is intrinsic to the emission from this region,
and not due to instrumental artefacts.
Lower surface-brightness emission is now also clearly detected,
enshrouding the jet all along its length. Due to this emission, it is
not possible to determine from the image whether the seeming ``gap''
is a true spatial break/decollimation of the jet, or whether the
emission from the main jet is just dimming and mixed with the surrounding emission in
this region (see also Sect.~\ref{sec:briprof}).

{\bf Arcs:}
In addition to the main spatial features described above, visual
inspection of the mosaic image at small scales around the pulsar reveals
the presence of an ``arc'' departing from the pulsar position towards the
counter-jet (see inset in Fig.~\ref{fig:j11}), with hints for the
presence of a second, almost symmetric, arc west of the pulsar
(i.e. towards the main jet).  Because of the relatively low S/N level
of the arc emission at larger distances from the pulsar, and to the
presence of diffuse emission around the PWN edge, it is, however,
impossible to verify whether this arc structure exists only between
the pulsar and the base of the counter-jet, or if it extends further
away.  The arcs can be followed up to $\sim$12\arcsec\ and 18\arcsec\
from the pulsar, in direction to the main jet and to the counter-jet,
respectively.

\begin{figure}
  \includegraphics[width=0.5\textwidth]{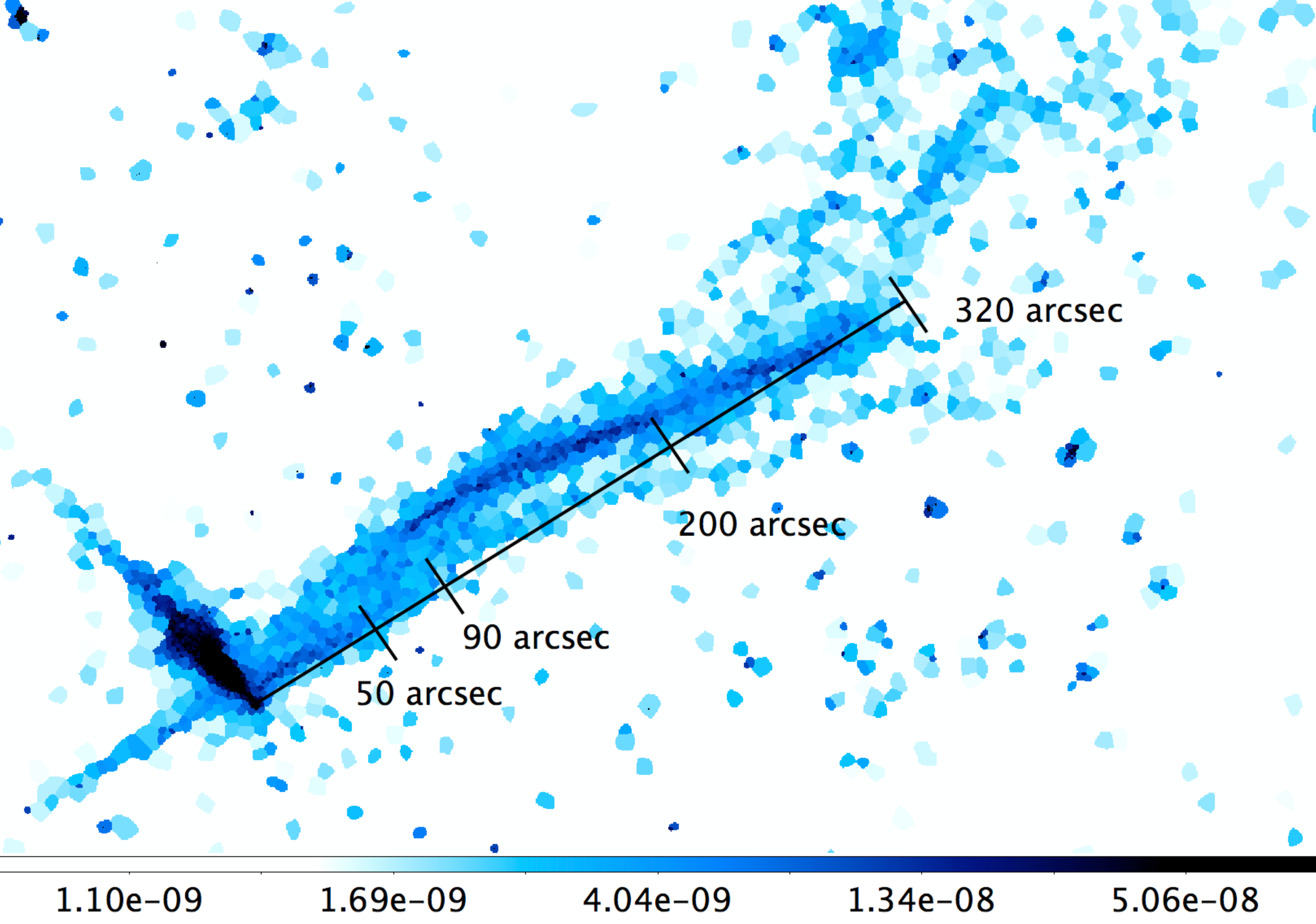}
  \caption{Mosaic image of \j11\ with Voronoi adaptive binning.
Each cell in this image has a S/N $\ge 3$.  The image is color coded in
units of photon/cm$^2$/s/arcmin$^2$ as detailed in the color map at the bottom of
the plot.  The pulsar, the main jet, the counter-jet, and the PWN are
all detected at high significance. The broader emission around the main jet
is also clearly detected.
The black ruler 
shows linear distances along the jet.}
  \label{fig:voronoi}
\end{figure}

To reveal the signal in low surface-brightness regions, we extracted
an adaptively binned image of the \lhn. In order to do this, we
applied the Voronoi tessellation algorithm developed by
\cite{cappellari}, as implemented in \citet{eckert2015}. We used a
target S/N of 3 for the algorithm, which allows us to visualize
statistically significant brightness variations across the nebula. The
resulting image is shown in Fig. \ref{fig:voronoi}.  All the known
structures of the \lhn\ (pulsar, PWN, main jet and counter-jet) are
highly significant, as well as the diffuse emission around the main
jet. This diffuse emission appears organised in ``stripes'' almost parallel to the main
jet and more prominent at its far end.  Closer to the pulsar, the
algorithm is not able to resolve the apparent parallel lines
(Fig.~\ref{fig:j11}), the emission appears as more diffuse.  This
broad region of emission encircling the main jet is analysed in more
detail in Sect.~\ref{sec:briprof} and \ref{sec:model}.

\subsection{Pulsar proper motion}
\label{sec:proper motion}

The absolute astrometric accuracy of 0.6\arcsec\ 
of \chan\ permits to perform sensitive proper
motion searches of bright point-like sources and in particular of isolated pulsars, by comparing the position
measured in observations separated by several years
\citep[see e.g.][]{auchettl2015, Van-Etten2012, motch2009}.
Moreover, the relative positional accuracy between different observations 
can be significantly improved with respect to the absolute
accuracy by removing systematic uncertainties that affect in the same
way the different observations (see the \chan\ documentation at
  \mbox{\url{http://cxc.harvard.edu/cal/ASPECT/celmon/}}).
  Here, we used our previous 50~ks observation and the new 250~ks
  \chan\ observation of \j11\ to search for proper motion of the pulsar (details
  of each observation are listed in Table~\ref{tab:observations}).  In
  each observation the pulsar is almost on-axis, although the telescope
  was operated in different configurations.  To study the pulsar and the
  surrounding region, we used event files reprocessed without the
  vfaint mode background cleaning, as we included the old observation
  performed in FAINT mode.  We used as reference frame obs.ID 16007,
  which has the longest exposure time (116~ks) and therefore the
  largest number of field point sources detected.  We then registered
  all other exposures by creating a list of common field point sources
  with detection significance above 4~$\sigma$ (within 8\arcmin\ from
  the pulsar, to avoid including too strong boresite effects) between
  each frame and the reference one, to minimize the observed
  displacements. The CIAO tool \texttt{wcs\_match} was used to
  accomplish this.  The number of common field sources between the
  reference frame (where we identified a total of 36 field sources
  satisfying the above criteria) and each other frame varied between 8
  and 20, depending on the frame exposure.  We corrected the reference
  positions for known proper motions of the field sources (we found 4
  field point sources with optical counterpart in USNO B-2, for which
  a significant proper motion is known).

  By comparing the residuals of the field source positions after
  registration of the frames, we found a relative positional accuracy
  of 0.2\arcsec\ between all observations.  We verified that inclusion
  or exclusion of the four sources with high proper motion does not
  affect the final positional accuracy.  No significant pulsar
  displacement has been detected between the different epochs (which
  are approximately 2 years apart), resulting in a pulsar proper motion
  $\mu_{\textrm{PSR}} \le 0.3\arcsec/yr$ (at 3 $\sigma$ level).  This
  upperlimit is not constraining, it is consistent with the expected
  value of 0.03\arcsec/yr
  $ (v_{\textrm{PSR}}/ 1000\ \textrm{km/s}) \cdot (7\
  \textrm{kpc}/d_{\textrm{PSR}})$,
  assuming a pulsar velocity of 1000~km/s \citepalias{pavan13}.

\subsection{Brightness Profiles}
\label{sec:briprof}

\begin{figure}[!tbp]
  \centering
  \hspace{10pt} \includegraphics[width=0.32\textwidth]{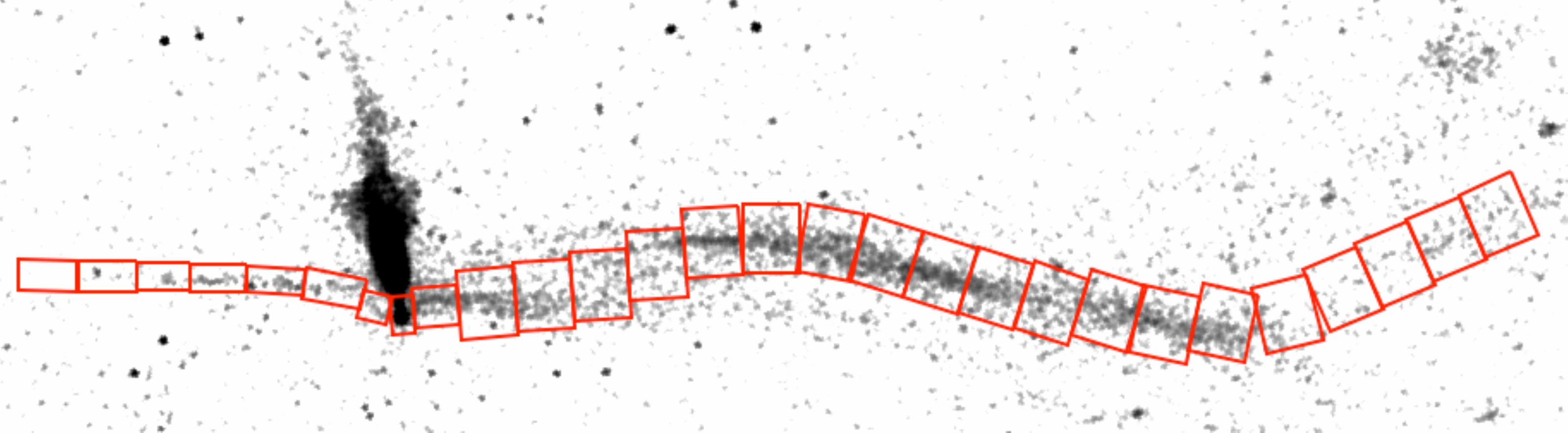}\\
  \includegraphics[width=0.45\textwidth]{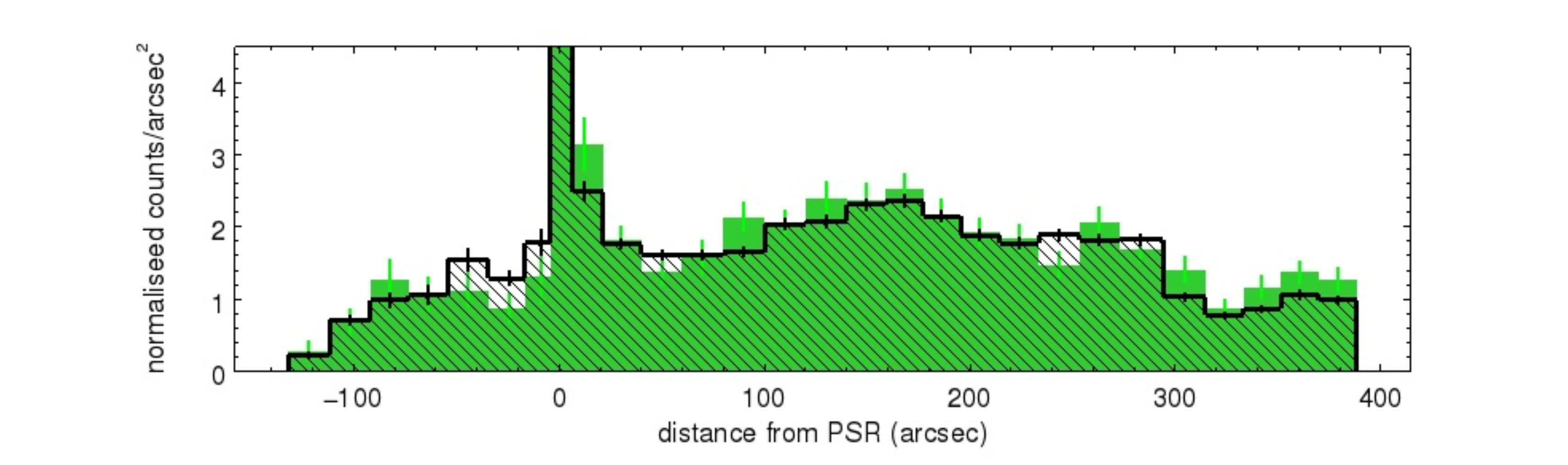}
  \\
\vspace{1.5cm}
  \hspace{3.5pt} \includegraphics[width=0.34\textwidth]{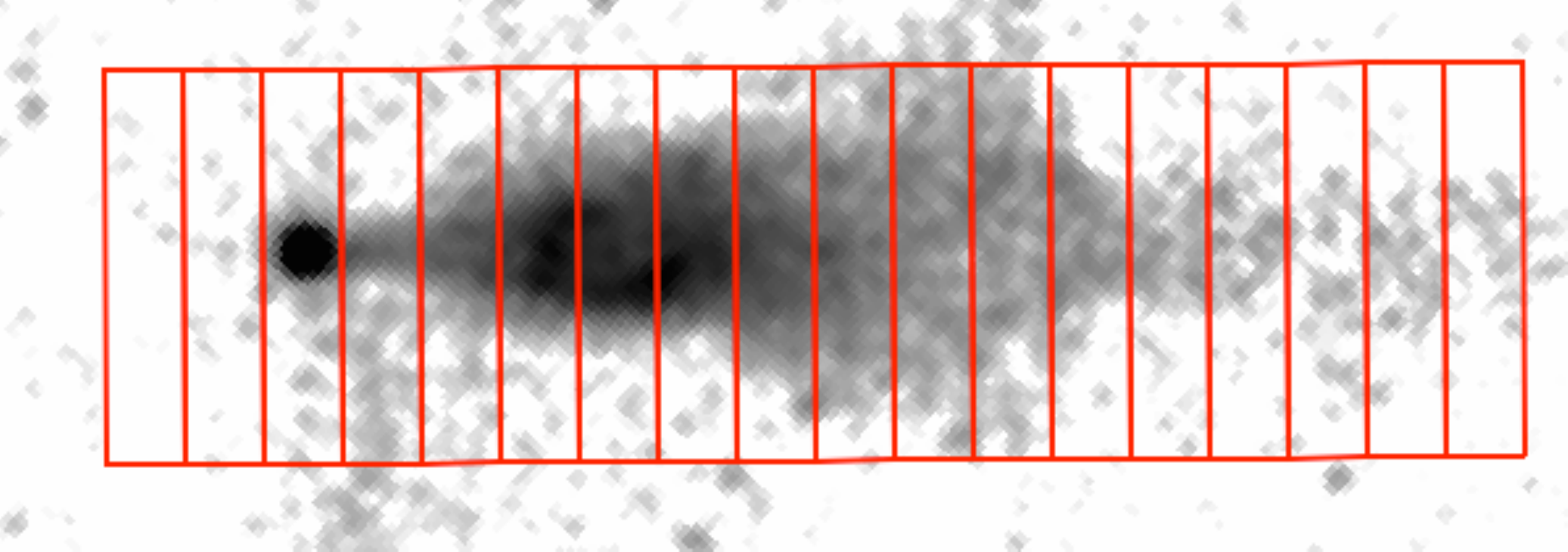}
  \includegraphics[width=0.45\textwidth]{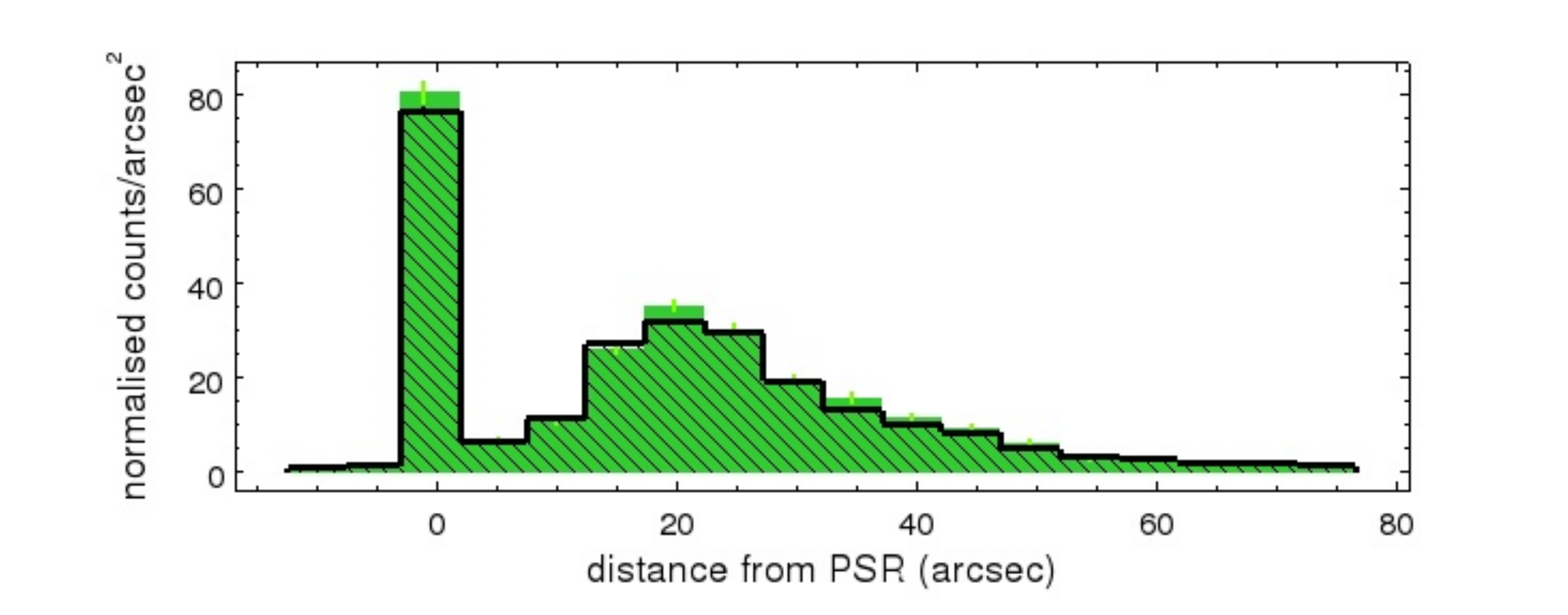}
  \caption{ Top panels: brightness profile as a function of distance
    from the pulsar, extracted along the
    jets (the brightness in the pulsar bin is not to scale), from the regions shown in
    the image above. The horizontal scale is the same in the profile plot and in the
    extraction region image.
    The black profile is extracted from the new 250~ks data, for
    comparison the profile obtained from the previous 50~ks
    observation is drawn as a filled green histogram.
    Bottom panels: same as above, but for extraction regions
    along the PWN.}
  \label{fig:brightness profiles}
\end{figure}

We extracted exposure-corrected, background-subtracted brightness
profiles along the PWN and the main jet, using counts in rectangular
regions.  We defined 20 regions along the jet, extended for 20\arcsec\
each, and 18 regions of 5\arcsec\ length each along the PWN (see
Fig.~\ref{fig:brightness profiles}).  From the counts recorded in each
region we computed the brightness profiles with the CIAO tool
\texttt{dmextract}.
In Fig.~\ref{fig:brightness profiles} we compare the profiles
extracted from the new 250~ks data with those obtained from the same
regions in the old 50~ks observation. 
The images are normalised for the corresponding exposure maps, thus
correcting for the effects related to the different integration time
and satellite roll angle (and hence also the different positions of the chips gaps
on the images).
The brightness profiles do not show any relevant difference between
the old and new observation (to within the uncertainties).  At
50-90\arcsec\ from the pulsar, where the images seem to show a spatial
break, we do actually not measure any significant brightness decrease
with this choice of the profile extraction regions.

\begin{figure}[!tbp]
\hspace{-1cm}
  \includegraphics[width=0.6\textwidth]{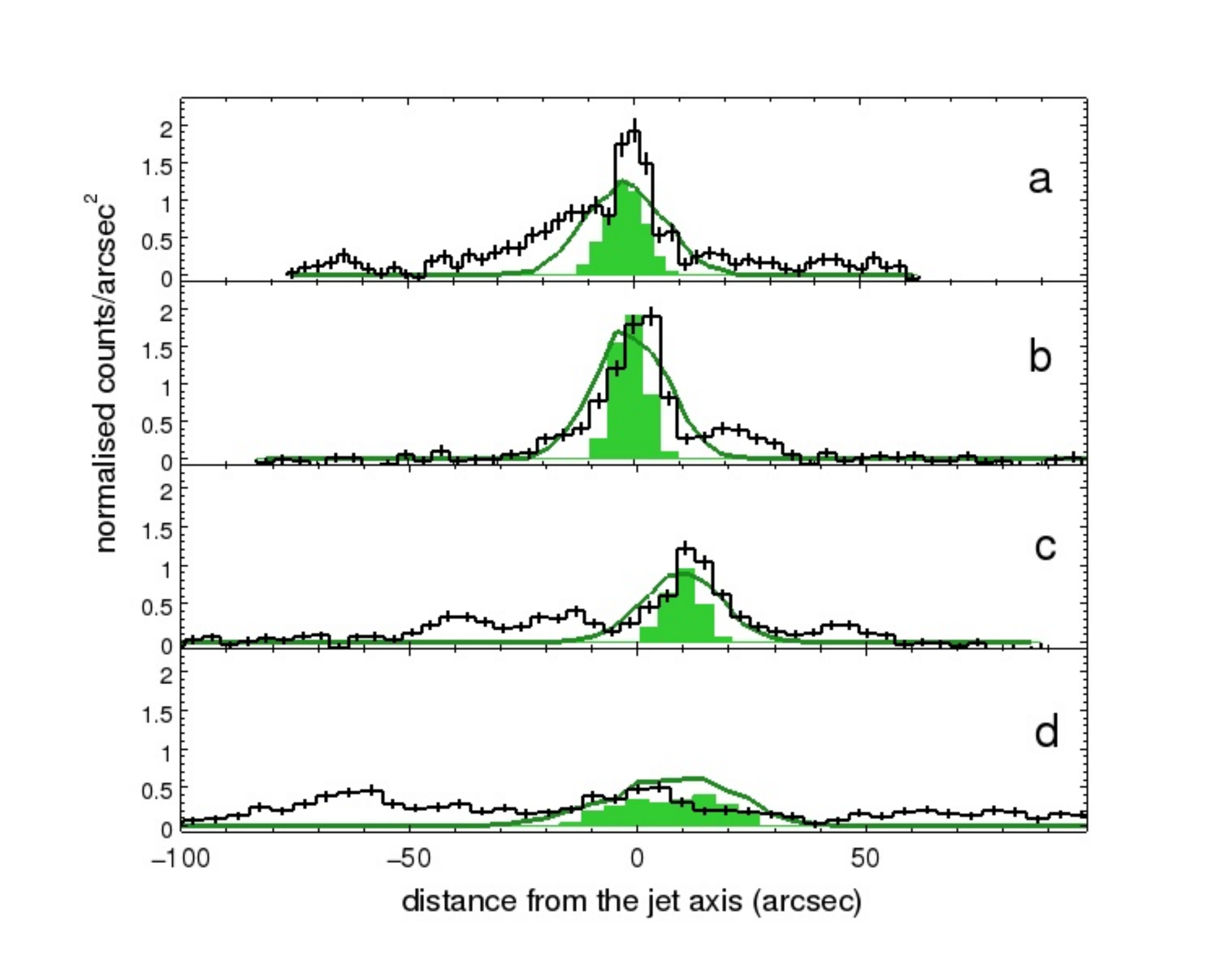}
  \caption{Surface brightness profiles (black histograms) extracted in
    four cuts perpendicular to the main jet (see regions 
    ``a'' -- ``d''  in Fig.~\ref{fig:j11}). We also plot for comparison the
    corresponding profiles obtained from a model with a single narrow
    (green histogram) or large (green solid curve) helical jet (see
    Sect.~\ref{sec:model}).  In each plot the distance from
    the jet axis increases towards the pulsar direction of motion (SW).}
  \label{fig:transv profiles}
\end{figure}

In our deep mosaic (Fig.~\ref{fig:j11}), an additional diffused
emission component is visible around the main jet. The Voronoi-binned
image (Fig.~\ref{fig:voronoi}) suggests that this emission comprises
several ``stripes'' developing almost parallel to the main jet. To
further investigate the structure of the region around the main jet,
we computed the brightness profiles in different cuts perpendicular to
the jet (Fig.~\ref{fig:transv profiles}).  These profiles clearly
highlight the presence of asymmetric X-ray emitting structures located
both north and south of the main jet. Given the relatively low S/N of
the data, we are currently unable to put firm constraints on the
geometrical shape of these structures (but we speculate on
possible modelling in Sect.~\ref{sec:model}).

\subsection{Spectra}
\label{sec:spectra}

We extracted from each \chan\ observation in
Table~\ref{tab:observations} all the spectra that are described in
Sect.~\ref{sec:mean-spectra} and \ref{sec:spectra-in-space},
by using the CIAO tool \texttt{specextract}, computing
at the same time the ancillary response file (ARF) and the response
matrix file (RMF) of each dataset. We used the same extraction
regions in all frames, and then combined the spectra of the
same region with the CIAO tool \texttt{combine\_spectra}. This allowed
us to obtain
merged spectra with higher S/N and their corresponding weighted ARF and RMF.
We verified that using the combined spectrum or fitting simultaneously
the spectra from the various Obs-ID provides the same results within
the statistical uncertainties.  In the following we thus discuss only 
the merged spectra.

\subsubsection{Average spectra of the extended regions}
\label{sec:mean-spectra}

We extracted the average spectra of the pulsar, the PWN and the jets from the
regions defined in Fig.~\ref{fig:j11}.  The extraction regions have been chosen
to avoid the chips gaps in each exposure where possible (the only
exception being the
counter-jet and the PWN, which
fall between two chips in all Obs-IDs).\\
All spectra are well described by a simple absorbed powerlaw model
(reduced \chisq = 0.85--0.99).  The results of the spectral analysis
obtained from the fits to the pulsar, the PWN and the main jet are
compatible with those reported previously in
\citetalias[see ][]{pavan13}.  
Also the spectrum of the counter-jet is well fitted
 (reduced \chisq = 0.9) with a simple absorbed powerlaw model.
  The best fit parameters are shown
in Table~\ref{tab:chanspec}.
All spectra are interpreted as synchrotron emission from relativistic
electrons, following the discussion in \citetalias{pavan13}.

\begin{table*}
  \caption{Best fit spectral parameters for the \lhn\ components.
    All spectra were fit using an absorbed powerlaw model (photon index $\Gamma$).  
    Uncertainties 
    are at 90\% c.l. on the spectral parameters and 68\% c.l. on the
    fluxes.}
  \center
  \begin{tabular}{@{}ccccc@{}}
    \hline
    \hline
    \noalign{\smallskip} 
    &  $N_{\rm H}$ & $\Gamma$ & $F_{\rm 2-10~keV}$ & $\tilde\chi^2$ /d.o.f.  \\
    &  {\small (10$^{22}$~cm$^{-2}$) }& & {\small (10$^{-13}$\ferg)} & \\
    \noalign{\smallskip} 
    \hline        
    \noalign{\smallskip} 
    Pulsar & 0.88$\pm$0.09  & 1.08$\pm$0.08 & 6.2$\pm0.15$  &0.85 / 175 \\
    \noalign{\smallskip} 
    PWN & 0.99$\pm$0.05  & 2.22 $\pm$0.06 & 6.1$\pm 0.1$  & 0.87 / 259 \\
    \noalign{\smallskip} 
    Main jet & 0.9$\pm$0.1  & 1.7$\pm$0.1 & 6.5$\pm 0.2$  & 0.99 / 233  \\
    \noalign{\smallskip} 
    counter-jet & 0.7$^{+0.4}_{-0.5}$  & 1.9$^{+0.5}_{-0.6}$ & 0.17$^{+0.02}_{-0.04}$ & 0.9 /12  \\
    \noalign{\smallskip} 
    \hline
  \end{tabular}
  \label{tab:chanspec}
\end{table*}

\subsubsection{Spatially-resolved spectra}
\label{sec:spectra-in-space}

\begin{figure}[!htbp]
  \centering
  \includegraphics[width=0.45\textwidth]{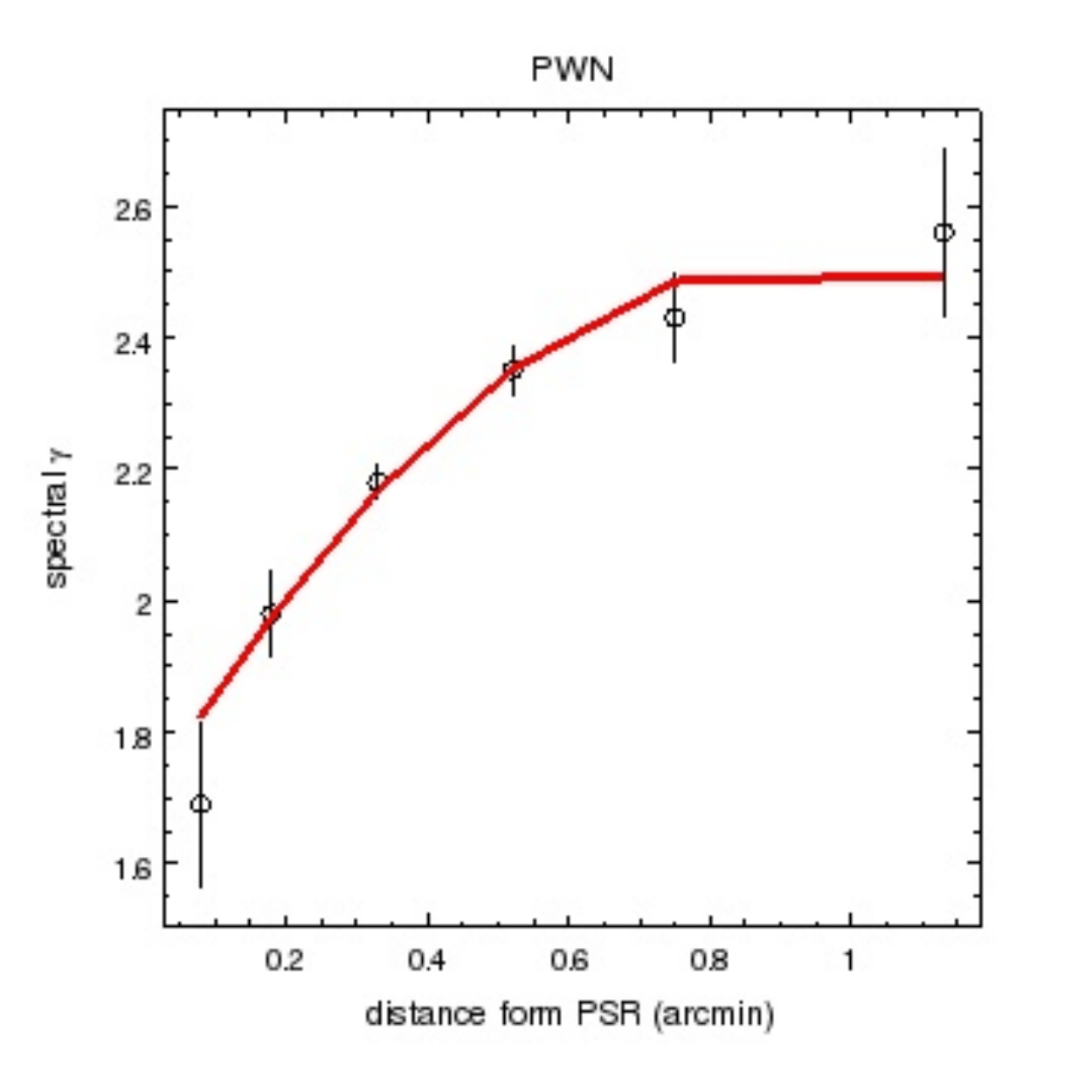}
  \caption{Photon index along the PWN (uncertainties at 1$\sigma$
    level). The best fit with a parabolic function is shown in red.}
  \label{fig:softening}
\end{figure}

In order to analyse possible spectral variations as a function of
distance from the pulsar, we extracted a number of spectra from small
rectangular regions covering the PWN (as shown in Fig.~\ref{fig:j11}).
The regions follow the brighter part of the PWN, to maximise the S/N
ratio.  We used the same background regions as defined above.  All
these spectra could be well fit with a simple absorbed powerlaw model.
The absorption column density remained constant within
uncertainties. We then fixed the column density $N_H$ among all
regions of the PWN to the average value of $9.9\,\times 10^{21}$
cm$^{-2}$. The photon index distribution is shown in
Fig.~\ref{fig:softening}.
The photon index of the PWN X-ray emission increases noticeably with
the distance from the pulsar.
A fit with a parabolic
function gives an acceptable description of the data (reduced
\chisq=0.7) and provides a significant improvement with respect to
the fit with a  
linear function (reduced \chisq\ =2.6).

\begin{figure}[!htbp]
  \centering
  \includegraphics[width=0.5\textwidth]{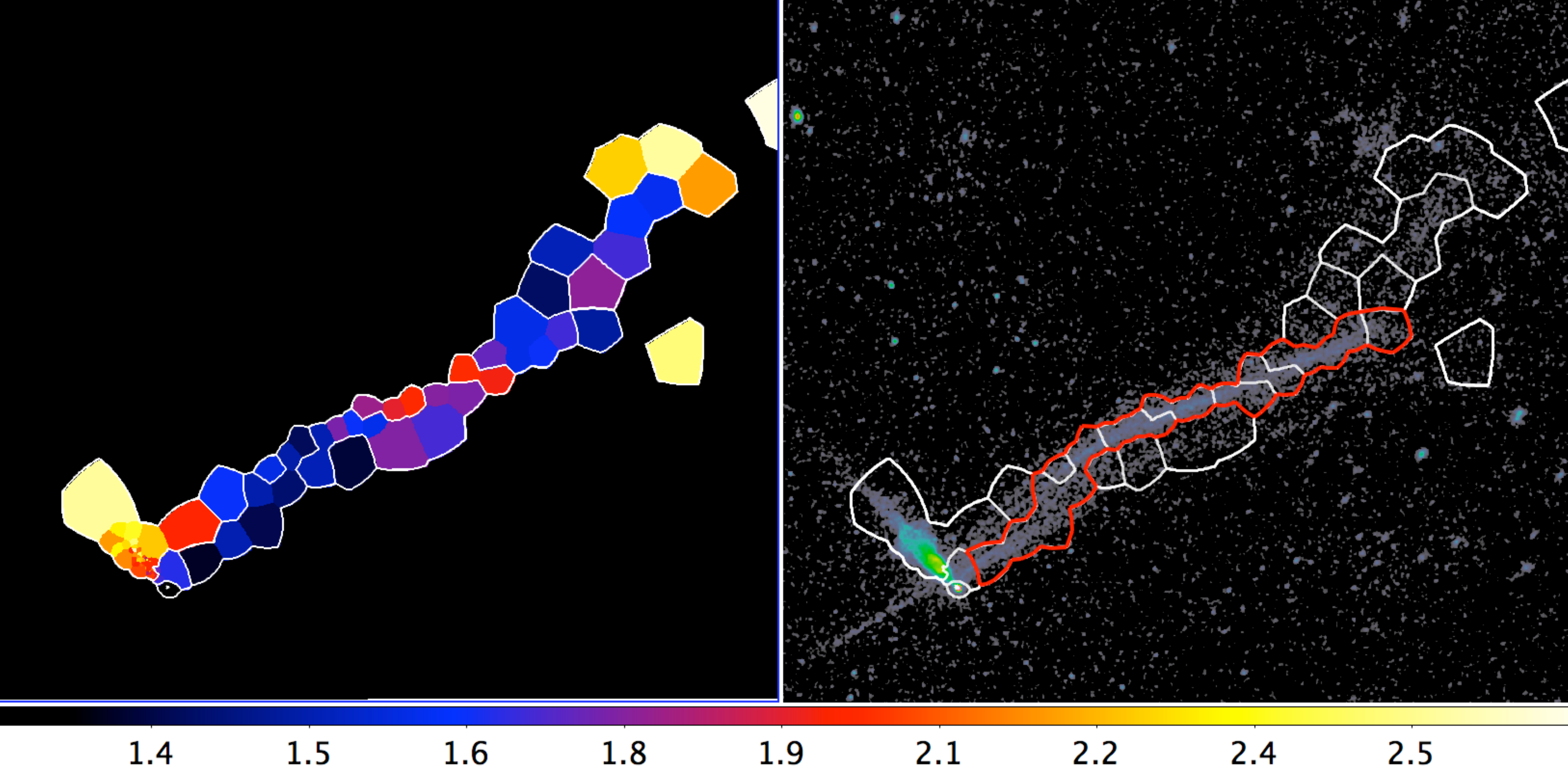}
  \caption{Left panel: Photon index map along the main jet and PWN.
    The colors represent different spectral indices, as shown in
    the colorbar at the bottom of the plot. Contours are drawn around
    the regions for clarity, at levels of
    $\Gamma = 1.1, 1.5 , 1.8, 2.8$.  Typical uncertainties on the
    spectral indices are on the order of $\pm 0.1$.  Right panel: the same
    contours are reported on the spatial map to aid the visual
    identification of the regions used for the spectral extractions,
    and of the regions clustering around similar values of $\Gamma$.
    The region in red was used to extract spectra from the main
    jet.}
 \label{fig:photon map}
\end{figure}

\begin{figure}[!htbp]
  \includegraphics[width=0.5\textwidth]{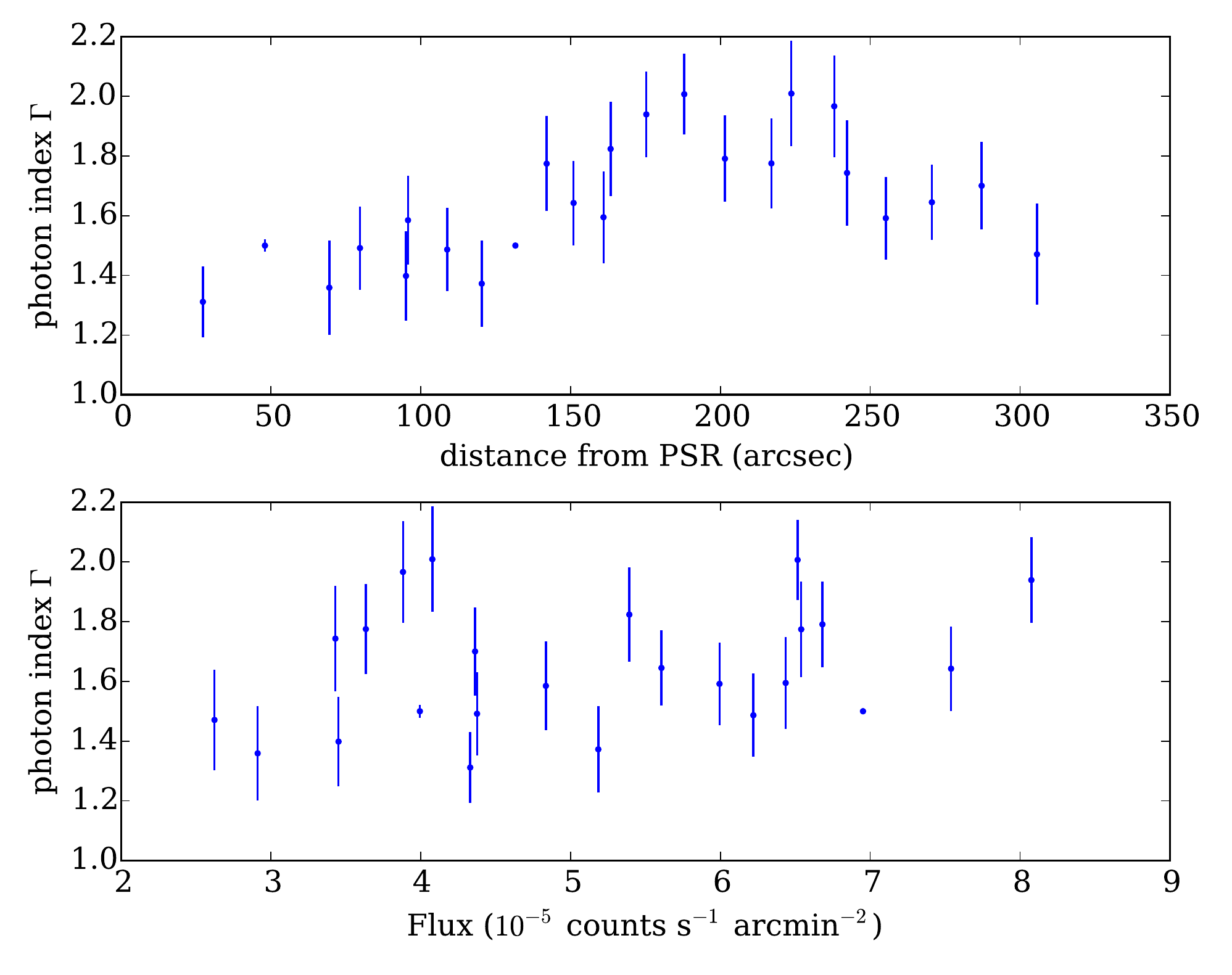} 
  \caption{Upper panel: photon index $\Gamma$ as function of
    distance along the collimated main jet (see extraction region in
    red in Fig.~\ref{fig:photon map}).
  Bottom panel: as for the upper panel, but here plotted
  against flux in each Voronoi cell. }
 \label{fig:correlation}
\end{figure}

We performed a similar analysis of photon index along the main jet. In
this case, however, a more complex variation of the photon index with
the distance from the pulsar is found. To study these variations in more
details, we extracted a photon index map (Fig.~\ref{fig:photon map})
using the method presented in \citet{rossetti}. We defined an adaptive
2D binning using the Voronoi tessellation technique presented in
Sect.~\ref{sec:imaging} with a target of 300 counts per bin. We then
extracted the surface brightness in five logarithmically-spaced energy
bands spanning the 1-6 keV range, where the main jet is predominantly
emitting. We estimated the local background rate in each band and
subtracted it from the data, adding in quadrature the uncertainty in
the background rate. We then used XSPEC to fold the model spectrum
with the \chan\ response files to create a template of the expected
count rate per band as a function of the photon index. The absorption
column density was fixed to its mean value of $N_H=9\times10^{21}$
cm$^{-2}$. A $\chi^2$ minimisation procedure was then used to fit the
templates to the spectra in each individual band and estimate the
photon index with its
uncertainty.\\
The photon index map confirms the smooth
softening along the PWN, and the complex
evolution of $\Gamma$ within the main jet. 
To further characterise the spectral profile along the jet, we
analysed first the correlation between $\Gamma$ and the distance, and
then between $\Gamma$ and the flux. This analysis was repeated for the collimated main jet
alone (red region in fig.~\ref{fig:photon map}), and for the brighter emission around it.
In the latter case we did not find any significant correlation (the
Spearman rank was $R_{S_{\textrm{Dist}}} = 0.3 \pm 0.1$ and
$R_{S_{\textrm{Flux}}}= 0.4 \pm 0.1$, respectively for the
$\Gamma$-distance and for the $\Gamma$-flux relations).
When considering only the collimated main jet, we found that $\Gamma$
follows a curved trend as a function of distance (see
Fig.~\ref{fig:correlation}).  As the Spearman rank can not be used in
the case of non-linear relationships, we split the sample into two
groups: the linear correlation rank is
$R_{S_{\textrm{Dist}}} = 0.9 \pm 0.1$ up to 200\arcsec\ from the pulsar, and
$R_{S_{\textrm{Dist}}} = -0.8 \pm 0.2$ between 200\arcsec\ and
320\arcsec, showing therefore a good correlation in the first portion
of the main jet, and a slightly less significant
anticorrelation in the farther part.
Following the same separation up to 200\arcsec\ and 320\arcsec (again along
the main jet only), we obtain for the $\Gamma$ vs. flux correlation:
$R_{S_{\textrm{Flux}}}= 0.7 \pm 0.1$ up to 200\arcsec\ from the pulsar,
while no correlation is seen in the range 200\arcsec-320\arcsec\ from
the pulsar ($R_{S_{\textrm{Flux}}}= 0.1 \pm 0.3$).

\subsection{Jet model}
\label{sec:model}

\begin{figure}[!tbp]
\centering
  \includegraphics[width=0.5\textwidth]{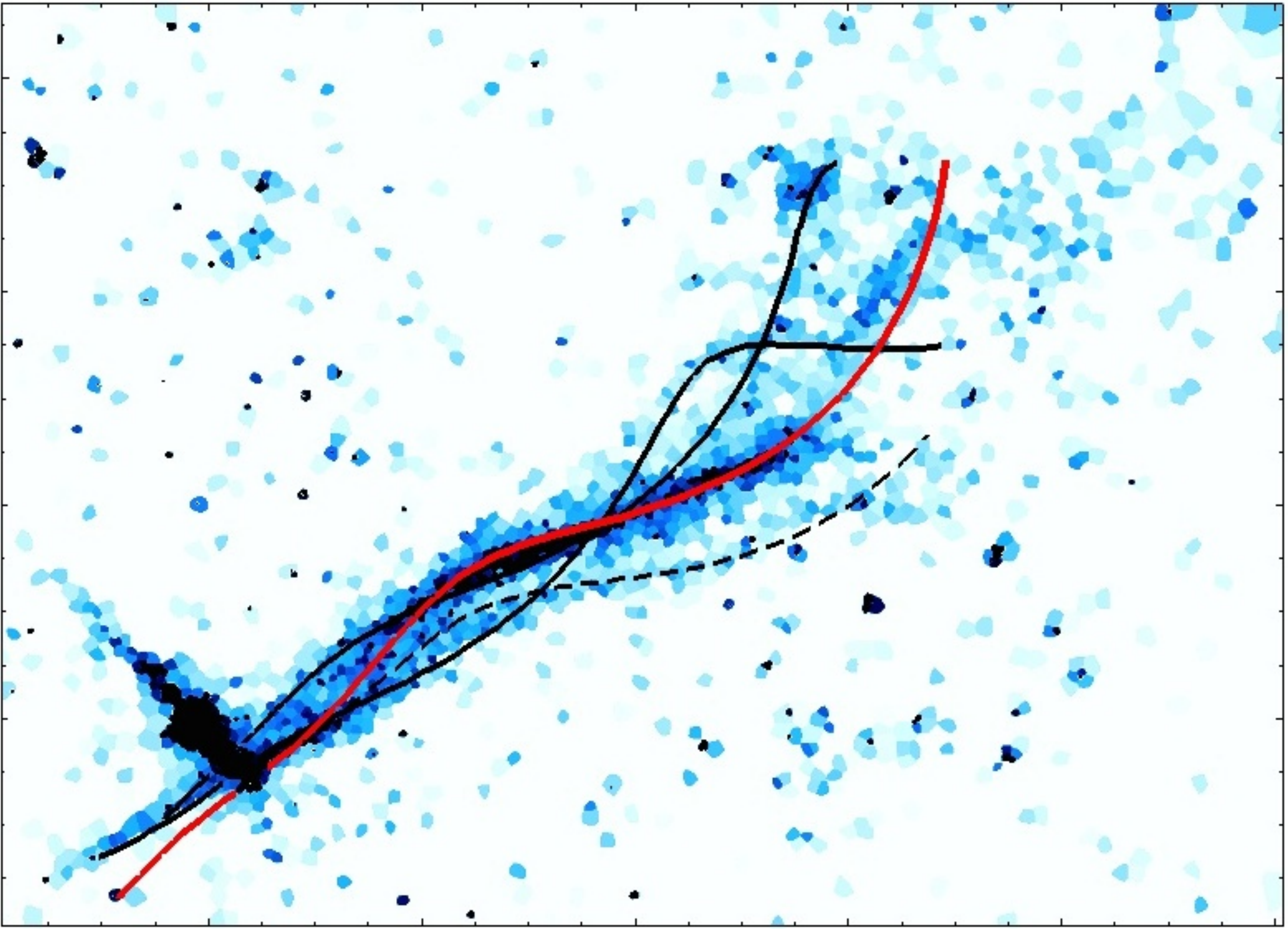}
  \caption{Solid red line: best fit to the main jet with a simple
    helical model, superimposed to the 250~ks \chan\ mosaic with
    Voronoi adaptive binning (Fig.~\ref{fig:voronoi}). The model is in
    relatively good agreement with the shape of the jet at distances
    $>$200\arcsec\ from the pulsar, however it fails to reproduce the
    regions closer to the pulsar, and the counter-jet.  In solid black
    lines are tentative additional helices, each with a different
    helical phase. A further tentative helix with different viewing
    angle is drawn with a dashed line.}
 \label{fig:helix}
\end{figure}

We applied the same 2D helical model described in \citetalias{pavan13}
to the new 250~ks image, to verify whether the helical pattern that we
detected in the previous shorter \chan\ observation can still fully
account for this deeper observation.  This is relevant in particular
for the portion of the jet within the first 100\arcsec\ from the pulsar,
where the new data set confirmed the loss of a coherent line structure
(see discussion in Sect.~\ref{sec:briprof}).

We used the \texttt{ciao/sherpa} tools and Python environment, and
used the \texttt{cstat} implementation of the Cash
statistic\footnote{In \texttt{ciao/sherpa}, the \texttt{cstat}
  implementation provides an approximate goodness of fit, with the
  reduced statistic resulting of order 1 for good fits.}.
We fit the helical model to the mosaic image, using the corresponding
exposure map, and adding a constant 2D function to simulate the
background level. In our model we considered a fixed value for the
particle bulk velocity $\beta = 0.8$ \citepalias[this value producing a
good agreement for the observed ratio between the mean fluxes of the
main jet and counter-jet, in the assumption that the two are
intrinsically identical and the observed difference is only due to
Doppler effects;][]{pavan13}.
We restricted the modelling to the main jet, excluding counter-jet and
the small scale arcs close to the pulsar.

The model includes a parameter used to modify the cross section of the helix
around its axis (forming a 2D gaussian cross section).  Given the presence
of significant broad emission as discussed above
(Sect.~\ref{sec:imaging}, \ref{sec:briprof},
\ref{sec:spectra-in-space}), we tuned this parameter a first time to
match the brighter collimated main jet, and in a second case leaving
the parameter free, to mimic the broader emission encompassing the
main jet.

In all cases we found that the fit statistic significantly improved
from a reduced value of 2.3 when only the background was included, to
a value of 1.5 (for the helical model parameters: inclination angle
38$\deg$, helical period 60~yr, cone semi-aperture angle of 4.5$\deg$;
for the full description of the model parameters see
\citetalias{pavan13}).  The residual map however shows that the image
is not well reconstructed, with only the large scale features being
reproduced.  Limiting the fitting to the regions at distances greater
than 90\arcsec\ from the pulsar, to avoid the region of the spatial
``gap'' closer to the pulsar, does not provide a significant improvement
in the fits results. The portion of the main jet closer to the pulsar
fails to be fully reproduced by the same model parameters found for
the farther part. If we include a helix symmetrical to the main one
to model the counter-jet, this further strongly departs from the
observed counter-jet direction (see the red line in
Fig.~\ref{fig:helix}).  Comparison of the brightness profiles
extracted from transversal cuts to the main jet (Fig.~\ref{fig:transv
  profiles}) shows that the single helix model does not follow
sufficiently well the profiles measured from the data: the narrow
helical model reproduces only in part the brightness peaks of the
collimated main jet; and a broad helical model fails to reproduce the
profiles observed from the extended emission around the jet.

Motivated by these mismatches of the single helix model, and by the
presence of significant broad emission distributed non-symmetrically
around the main jet (see transversal cuts in
Figs.~\ref{fig:voronoi} and \ref{fig:transv profiles}), we tentatively
included additional helical outflows in the model (see
Fig.~\ref{fig:helix}).  Given the relatively low surface brightness of this
region, we did not attempt to fit this model to the data. We inspected
only qualitatively whether this phenomenological model could
improve the description of the broad emission encompassing the jet,
with respect to the single helical model.

All the parameters describing the additional helices are fixed to
match the ones used for the main helix, except for the helical phase
and normalisation.  Whereas we found that the main collimated jet was
better described by an helix at phase\footnote{the helical phase angle
  is computed from an arbitrary zero point} 260 deg, we found that
the addition of further helices at phases 340 and 85 deg could
mimic quite satisfactorily the extended emission seen around the jet
(without attempting a fit, we computed the corresponding statistic
values with the ciao tool \texttt{calc\_stat}\ obtaining a reduced
statistic of 1.4 for 15350 degrees of freedom).  In addition a dim
fourth helix with the same phase as the dominant one (260$\deg$) and
slightly modified inclination angle of 30$\deg$ could reproduce the
emission observed south of the main jet at distances larger than
200\arcsec\ from the pulsar (see Fig.~\ref{fig:voronoi}).
Together with the helical phase and normalisation, the three
additional helices were also slightly shifted to match the main
features of the broader emission. As a side result, the launching
points of the additional helices come out to be aligned with the PWN
axis, backwards with respect to the dominant helix.

\section{VLT/FORS2 H$\alpha$ images}
\label{sec:vlt}

Our group obtained a VLT imaging observing run (run 092.D-0729, PI: Pavan),
carried out in service mode with FORS2 (\citealt{Appenzeller1998}) at
Cerro Paranal on December 23th, 2013. The aim of the observations was
to search for a bow-shock created by the supersonic motion of the pulsar
in the ISM.

The observations were performed using a narrow-band filter centered on
the H$\alpha$ line.  The total exposure time in this filter was 1 hour
36 minutes.  Additional short exposures (T$_{exp}$=35s) were also taken
with the \textsc{R\_SPECIAL} filter in order to have a handle on the
continuum emission.

The images were reduced using the \textsc{Theli} pipeline (\citealt{Schirmer2013,Erben2005})
which takes care of all processing steps, including bias subtraction, flat fielding, astrometric calibration
and coaddition. 

The H$\alpha$ narrow-band images were calibrated by observing the
spectrophotometric standard LTT4364 (\citealt{Hamuy1992,Hamuy1994})
with exposure times short enough to avoid saturation of the
detector. These images were reduced in the same way as the science
images. The tabulated flux of the standard star $f(\lambda)$ was
integrated within the filter transmission $R_{H\alpha}(\lambda)$ and
this value was compared with the measured count rate $c_{H\alpha}$.
The derived zeropoint following this procedure is

\begin{equation}
k = \frac{\int  f(\lambda)  R_{H\alpha}(\lambda) }{ c_{H\alpha}
  \text{[counts/s]} }
= 3.97 \times 10^{-18}$\,ergs\,s$^{-1}$\,cm$^{-2}$\,counts$^{-1}
\end{equation}

\begin{figure}[!htbp]
  \includegraphics[width=0.5\textwidth]{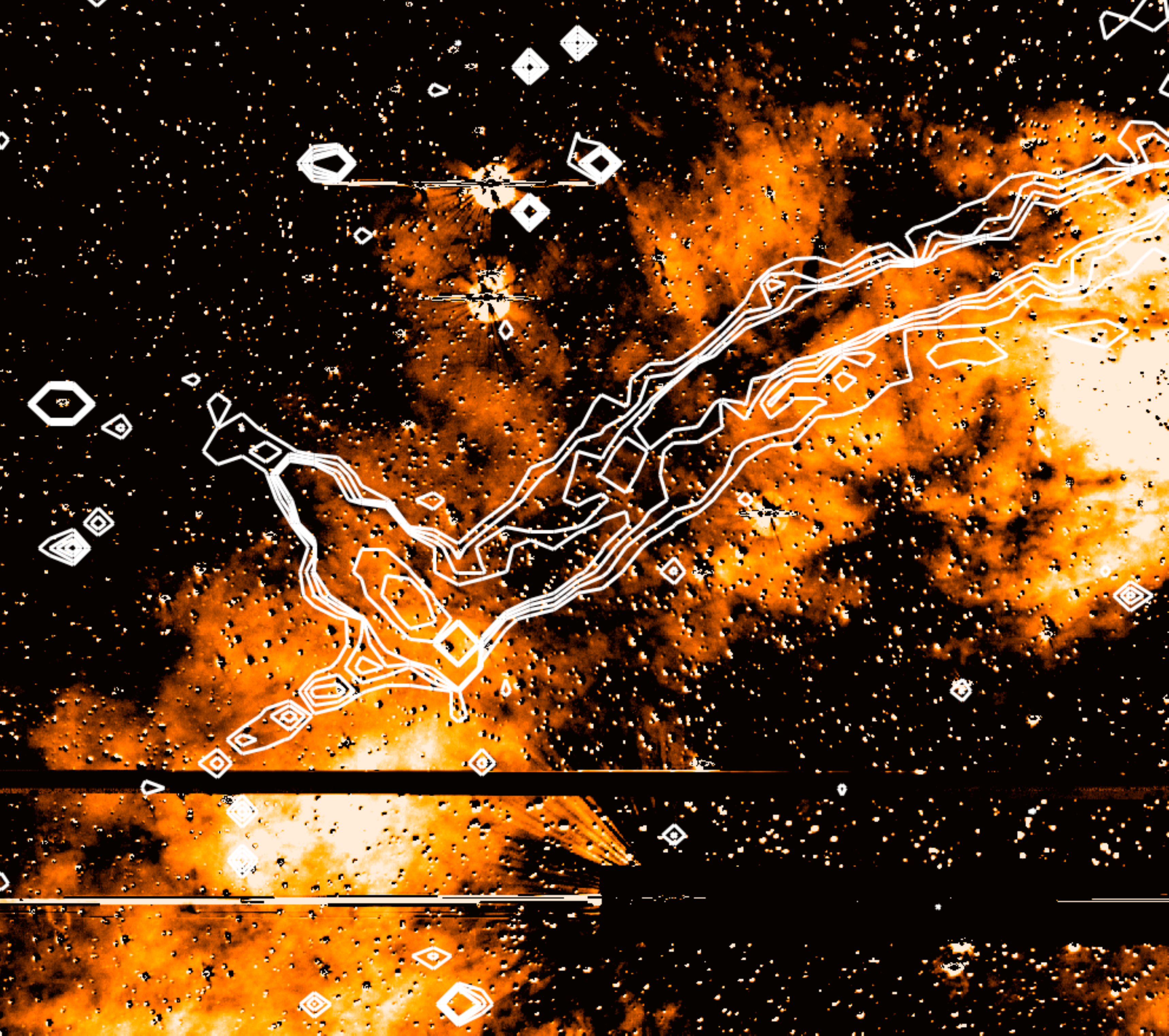}
  \caption{The \lhn\ region in \Halpha\ narrow band (continuum
    subtracted). Overplotted are the X-ray 
    contours from the mosaic image in Fig.~\ref{fig:j11}.
    At the bottom of PSR~J1101-6101 a bright field star was masked with a MOS occulting bar.
    The diffuse emission is likely unrelated to the \lhn, as discussed in the
    text, and prevents detection of any possible emission from a
    bow-shock in front of PSR~J1101-6101.}
 \label{fig:Halpha}
\end{figure}

Visual inspection of the narrow-band and continuum subtracted images
(Fig.~\ref{fig:Halpha}) reveals a general nebulosity over a large
fraction of the field-of-view.  This nebulosity together with the
general crowding of stars, typical of the galactic plane, makes the
detection of the bow-shock difficult.  We computed upper limits on
the bow-shock emission by placing 100 random apertures of
1$\times$1\arcsec$^2$ around the expected region of the bow shock,
obtaining
\mbox{$F(H\alpha)_{BS} \le 1.25 \times
  10^{-17}$\,ergs\,s$^{-1}$\,cm$^{-2}$\, arcsec$^{-2}$} at the 3$\sigma$ level.

\section{Discussion}
\label{sec:discussion}

\subsection{Main jet}

The new deep \chan\ observation was optimised to avoid any
instrumental feature falling along the main jet (chips gaps, dead
columns). The mosaic image that we obtained (Fig.~\ref{fig:j11})
confirmed the large scale spatial modulation of the main jet, as
already detected in the previous \chan\ observation
\citepalias{pavan13}, as well as the presence of a spatial
discontinuity along the main jet, between 50\arcsec\ and 90\arcsec\
from the pulsar. A significant broad emission surrounding the main jet is
now also detected (see e.g. Fig.~\ref{fig:voronoi}
and Fig.~\ref{fig:transv profiles}).
Whereas in \citetalias{pavan13} we found a good
match between the main jet and a simple helical model, in our new
images the same model shows some discrepancies with
the data. The helical modulation seem to be still
a good first order approximation, but
the higher statistics now available shows small scale departures from
such a relatively simple picture.
As discussed in \citetalias{pavan13}, a helical pattern along a
jet could be due either to a modulation of the jet launch direction by
a free precession of the pulsar or to the
development of kink instabilities along the jet \citep[see
e.g.][]{Lyubarskii1999, mollthesis2010}.  In this latter case it seems natural to
expect some small scale departures from an overall helicoidal trend.
In particular, the spatial discontinuity of the main jet of \j11\ might
resemble the filament breaks described in laboratory plasma jets
where macroscopic kink instabilities are accompanied by microscopic
Rayleigh-Taylor instabilities occurring at particular regions along the
magnetised plasma jet \citep{moser2012}.  These second instabilities
can erode a portion of the plasma jet, either completely or partially,
depending on the relative scales between the smallest filament
diameter reached and the plasma ion skin depth level
\citep{moser2012}.  Whereas the lab jets clearly develop on different
scales with respect to the jet at hand here, the characteristics of
jets produced in the laboratory are expected, as highlighted by the
same authors, to be largely scale invariant and therefore applicable
to astrophysical jets as well.
The surface brightness profile extracted along the main jet of \j11\
(Fig.~\ref{fig:brightness profiles}) shows no significant decrease in
the region of this spatial ``gap''. The main jet in IGR J11014-6103
seems therefore to keep physically undisrupted. Under the kink
instabilities scenario, the ``gap'' could then be interpreted as a
region where the Rayleigh-Taylor instabilities would have failed to
completely erode the jet.  Alternatively, the gap region could be
interpreted as the projected superposition of the main jet and the
broader diffuse emission surrounding it, possibly due to several
outflows (see below), degrading the coherence seen otherwise along
other regions in the jet.  Conclusive evidences for intrinsic
decollimation of the jet due to Rayleigh-Taylor instabilities or the
superposition of several emission components to explain the diffuse
emission along the gap cannot be derived from the observations
reported here.

The photon index distribution along the broad region encompassing the
main jet is rather complex, and no clear trend has been observed for
the photon index neither as a function of distance from the pulsar, nor
as a function of flux intensity (see
Sect.~\ref{sec:spectra-in-space}).  However, when restricting the
analysis only to the brighter collimated main jet, we found a clear
correlation between the photon index $\Gamma$ and distance from the
pulsar (see discussion in Sect.~\ref{sec:spectra-in-space}), with
positive correlation up to 200\arcsec, followed by anti-correlation
between 200\arcsec\ and 320\arcsec.  The first linear positive
correlation between $\Gamma$ and distance could be interpreted as
softening of the emitting particles farther away from the pulsar, e.g., due to
synchrotron losses.The sudden change to an anticorrelation at larger
distances is however more difficult to explain in this scenario.

Keeping in mind the overall helicoidal modulation of the jet, we
investigated the possibility that the variations in the observed
photon index could be produced by the Doppler effect. 
When the synchrotron emission is produced by isotropically distributed
electrons embedded in a magnetic field, the spectral index is a
Lorentz invariant, i.e., if the intrinsic flux follows a power-law
distribution in frequency space, the observed flux will also follow a
power-law with the same index (see for example the discussions
in \citealp{Rybicki-Lightman}, 
see also \citealp{BBR} for extragalactic jets, and the treatment for
different jet dynamics in \citealp{lind-blandford}). If the spectrum is intrinsically
curved, however, some further complexity may arise \citep[see e.g.,
Fig. 5 in][for the case of a cutoff-power-law]{Fraix-Burnet1997}. For
an approaching source, the Doppler boosting shifts the emitted
frequencies towards higher observed frequencies in the observer
frame. The observed spectrum could then appear overall harder than the
emitted one, if the curvature in the emitted spectrum
falls in a suitable range of energies.

In the case of a helical jet, as for \j11, assuming that approaching
and receding regions intrinsically emit the same spectrum, one could
then expect to observe harder spectra in approaching regions, if
curvature or breaks are present in the original spectrum.  Such breaks
could be present in the data but possible be unresolved in the spectral
fit.  The total flux of the approaching regions would also be
increased due to Doppler boosting, therefore one should expect an
anticorrelation between photon index $\Gamma$ and flux.
Along the jet of \j11, however, the positive and negative
$\Gamma$-distance correlation makes it difficult to explain the
spectral modulation along the jet by Doppler boosting.

Alternatively, the complex photon index trend could arise from either
particle re-acceleration along the jet \citep[see,
e.g. ][]{rieger2007}, or from varying conditions of the underlying
magnetic field. In the latter case one would naively expect softening
of the emission in correspondence to regions of higher magnetic field,
where particles cool more efficiently.  These same regions would yield
also higher total fluxes, therefore providing a positive $\Gamma$-flux
correlation. This could explain the spectral trend up to 200 arcsec,
but not the hardening later on.

When applying the simple helical model discussed in
\citetalias{pavan13} to the jet-like features, we found that, while
the single helical model can still reproduce the overall morphology of
the jet, it fails to reproduce in particular the portion of the jet
within the first 100\arcsec\ from the pulsar and the counter-jet, both
for the collimated main jet, and for the broader emission region
around it (see Sect.~\ref{sec:model}).  Whereas the counter-jet
mismatch could be due to non completely symmetric outflows and polar
caps of the pulsar \citep[as seen already both in isolated and in
accreting pulsars, see e.g.][]{harding2011, bogdanov2014, venter2015}, we
noticed that the inclusion of several simultaneous helices with
different launching phases could recover qualitatively the counter-jet
direction. The different helices are naturally encompassing the main
jet towards north or south directions depending on the distance from
the pulsar, rather than being symmetric around the main jet.  Although no
fit of the multiple helices model was possible, due to the low surface
brightness of the diffuse emission, the comparison between the region
covered by the broad emission and the projection of the different
helices hints towards an improvement of the modelling of this region
with respect to a single broad helix.  The complexity of photon index
$\Gamma$ with respect to both distance and flux could be at least
partially relaxed in this case, as the broad emission would derive
from the projection of several independent helices, each with a
different projected $\Gamma$ distribution.
A side outcome of this model is that the launching point of the
additional helices turns out to be aligned with the PWN axis and
positioned backwards with respect to the dominant helix.  This
alignment may naively recall residual emission from the main jet,
launched at different epochs in the past.  However, in this ballistic
jet model the past helices
should be largely detached from the pulsar path, reflecting the travel
time of the pulsar between the different launch points. The broad
emission surrounding the main jet in \j11\ is instead detected down to
the launching points of the hypothetical additional helices.  The
tentative description of the broad emission encircling the main jet as
being due to multiple helices necessarily needs therefore to involve
the simultaneous presence of all of them.

A current hypothesis for the Crab jet foresees that the pulsar ``jets'' can
be launched through the collimation of shocked wind plasma by hoop
stresses in the wind magnetic field \citep{Lyubarsky2002}. This would
occur, in the Crab and other low velocity pulsars, in locations above the
pulsar polar caps. Obviously, in this scheme no simultaneous presence of
different helices is expected.
In the case of a relativistic pulsar wind and strong velocity shears
in the PWN we note, however, that the overall geometry and the
occurrence of hoop stresses could be modified \citep[see for
example discussion in][]{kargaltsev2015}.

An alternative interpretation for the main jet would be that of highly
energetic particles being accelerated at the termination shock and
then trapped by the underlying ISM magnetic field, as already proposed
(and still debated) for the Guitar nebula \citep{bandiera2008,
  johnson-wang-10, Hui:2012fk}.  The properties observed from the main
jet only (photon index distribution, geometrical pattern, brightness
profile) don't provide however definite clues on either this model or
the ballistic jet one. In the case of electrons diffusing into the
ambient magnetic field, the existence of additional helices would
require the presence of a very particular geometry of the ISM magnetic
field.  Also in this case, no trapping of particles by the ISM
magnetic field is supposed to occur at places other than the front
shock.  At odds with the ballistic jet scenario, however, in the
hypothesis of diffusion of particles in the surrounding magnetic field,
past helices could show no significant detachment from the
pulsar path, due to possible backwards diffusion of particles along
the field lines.
The ``jets'' would then be a
visualisation of the ISM magnetic field in the surroundings of \j11
(compare for example the ISM magnetic field lines in
\citealp{giacinti2013}; see also the discussions on the Double Helix
Nebula, \citealp{morris2006, torii2014}).
The apparent organisation into helices that
share the same parameters and differ only for the helical phase angle would
however pose severe difficulties to this scenario as well, in particular
requiring these conical helices to have all the same aperture angle
and to converge towards the pulsar location.
To relax this apparent conflict, we note that the motion of the pulsar in
the ISM could affect the distribution of the surrounding ISM material,
together with its magnetic field.

\subsection{PWN}
\label{sec:discuss:softening PWN}
The magnetic field $B_{\rm PWN}$ in the PWN of the \lhn\ was
estimated in \citetalias{pavan13}  assuming that a single electron
distribution was responsible for both the X-ray and radio emission
observed along the tail. The angular separation between the emission
peaks at these wavelengths, together with a flow velocity constrained
to be $v_{\rm flow} \geq v_{\rm PSR}$, where $v_{\rm PSR} \gtrsim
1000$~km/s is the pulsar velocity through the ISM, was used to derive
$B_{\rm PWN} \sim 10-20~\mu$G. The new \chan\ observations
reveal a significant softening of the X-ray spectrum along the PWN
axis. Close to the pulsar, the angular scales on which the variations are
observed are as small as $\sim
0.1\arcmin-0.2$\arcmin, which corresponds to a linear scale of 
$ \ell \sim 6 \times 10^{17}$(d$_\textrm{PSR}/7$~kpc)~cm.
Assuming that the spectral variations are due to
synchrotron 
losses of the electron population in the nebular magnetic field 
implies that the emitting population already cooled down while
travelling these distances.
Electrons mainly radiating in the $\sim$1-10~keV X-ray band require
energies
$E_{\rm e^{-}} \approx 45 \, (B_{\rm PWN}/ 100~\mu G) ^{-1/2} \,
(\epsilon/5~\rm{keV})^{1/2}$~TeV,
with $\epsilon$ the X-ray photon energy. Equaling the synchrotron
cooling time for these X-ray emitting electrons
$t_{\rm sync}(\epsilon) \approx 9 \times 10^{8}\, (B_{\rm PWN}/
100~\mu G) ^{-3/2} \, (\epsilon/5~\rm{keV})^{-1/2}$~s
to the dynamic time for the cooling electrons to propagate within the
nebula $\tau \sim \ell /v_{\rm PSR}$, an updated value for the
nebular magnetic field $B_{\rm PWN} \gtrsim 30~\mu$G is retrieved.

The X-ray spectral profile along the nebula axis displays 
a monotonic steepening, ranging from \mbox{$\Gamma$ = 1.7
  to 2.5} (see Fig.~\ref{fig:softening}), characteristic of
synchrotron cooling. The softening rate is however higher at regions
close to the pulsar, and flattens down at larger distances. Such profile
differs from what is expected in a simple scenario in which radiating
electrons are just advected away along the tail \citep[see, e.g., the
1D toy model applied to the PWN N157B in][]{Chen2006}. Similar
profiles for the X-ray photon index have also been observed in other PWNe,
e.g. 3C 58 \citep{Slane2004}, G21.5--0.9 \citep{Slane2000}, or MSH
15--52 \citep{An2014}, and have been interpreted in terms of
energy-dependent diffusion caused by Rayleigh-Taylor instabilities at
the boundaries of the PWN and/or due to instabilities in the nebular
magnetic field \citep{Tang_Chevalier-2012, Begelman1998}. An unstable
magnetic field structure deviating from a pure toroidal component may
indeed be expected in the strongly perturbed nebula of \j11\ given the
high proper motion velocity of the system through the ISM.

The PWN initially opens up forming a wider cone, and includes also a
more collimated tail visible up to 60\arcsec\ from the pulsar (as
described in Sect.~\ref{sec:imaging}). It could be compared to the
structures seen in other bow-shock PWNs, and noticeably in the
``Mushroom Nebula'' powered by PSR B0355+54 \citep{mushroomPWN}.  The
main differences between the two objects are the presence in the \lhn\
of a strong narrowing region immediately behind the pulsar (up to 7\arcsec\ from
the pulsar, see Fig.~\ref{fig:brightness profiles}), which is not seen in
the Mushroom; and the absence of extended emission surrounding
PSR~J1101-6101.  A displacement between the highest intensity of the
bow-shock PWN and the pulsar location has been observed in the case
e.g. of the Turtle nebula \citep[\morla; ][]{deluca2011, deluca2013,
  marelli2013}. In this case, however, also an alternative
interpretation was drawn, where the X-ray trail was modelled as
thermal emission from the shocked ISM along the pulsar path
\citep{marelli2013}.  In the case of \morla\ no \Halpha\ emission was
detected, since the surrounding ISM is fully ionised. Similarly, in the case of
PSR~J1101-6101, we could not detect \Halpha\ emission either (see
Sect.~\ref{sec:vlt}), although the non-detection in our case is not
conclusive given the presence of a large surrounding nebulosity.
In the case of the Mushroom nebula, the presence of a jet 
has been invoked to explain the more extended ``stem'' part of the
PWN.
For \j11, the rotation axis may be oriented in the direction of the
jet/counter-jet, so a new, roughly perpendicular outflow, which
moreover falls right at the dissecting line of the PWN, seems difficult to justify. 
The presence of two different
components in the PWN of \j11\ could also be explained by differences
in the magnetic field confinement, and by the presence of a strongly
relativistic component of the wind \citep{bucciantini2005}.

Besides the clear differences discussed above, we note that the X-ray
emitting PWN population is characterised by a large variety of
morphologies \citep[see e.g.][]{kargalstev-pavlov-2008,
  kargaltsev2015} and both the Mushroom nebula and the \lhn\ could be
explained under the same general bow-shock PWN model, assuming
differences in the ISM and pulsar properties (for example alignment
between the pulsar magnetic field axis
and its direction of motion, or spin axis).

\subsection{Arcs}
\label{sec:discuss:softening PWN}

Much closer to the pulsar, towards the
counter-jet, we observe in the \chan\ mosaic a well defined arc
structure (Fig.~\ref{fig:j11}). The image also suggests the presence of a symmetric arc
departing from the pulsar towards the main jet.  It is not clear whether
these structures directly connect smoothly to the jets, or whether
they continue in the direction of the PWN (see Sect.~\ref{sec:imaging}).
The ``arcs'' could be interpreted either as emission from the shocked ISM (a
velocity of 1000 km/s, as the ones inferred for PSR~J1101-6101, is
indeed considered a lower limit for a shock to start emitting X-rays)
or as outflows from the pulsar itself, in analogy to what is observed
already in several other PWNs \citep[see e.g. in Geminga,
][]{kargaltsev2015}.  

\subsection{ PSR-SNR association}
\label{sec:discuss:softening PWN}

\begin{figure}[!tbp]
  \centering
  \includegraphics[width=0.5\textwidth]{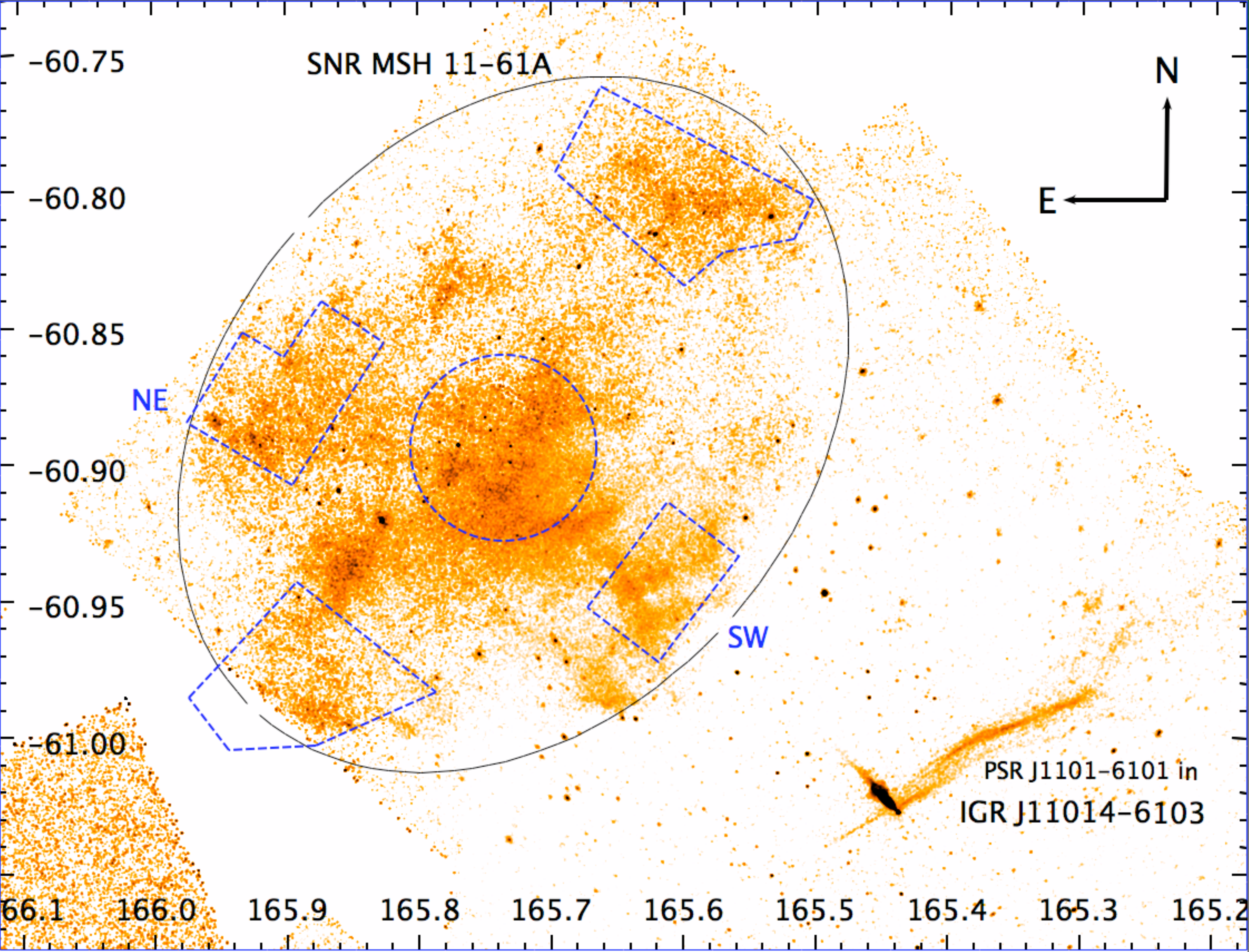}
  \caption{ Mosaic including all archival \chan\ observations of
    the \SNR\ and \j11\ region.
    Dashed blue polygons are reproduction of the extraction regions used by
    \citet{Garcia:2012fk} for spectral analysis in the SNR (see
    discussion in Sect.~\ref{sec:discussion}).}
  \label{fig:largeMosa}
\end{figure}

Current 3D simulations drawn in the ideal case of
non-rotating progenitor stars, predict that asymmetric ejections
caused by hydrodynamic instabilities could impart strong kicks to a
new born neutron star during a core collapse event \citep{Janka-2012,
  Wongwathanarat2013}.  
In this case, following \citet{Wongwathanarat2013}, heavy elements
should be distributed asymmetrically in the SNR, clustering in the
direction opposite to that of the pulsar kick.
To this aim we inspected the published results about a spatially
resolved spectral analysis of \SNR\ performed with XMM by
\citet{Garcia:2012fk} and with Suzaku by \citet{Kamitsukasa:2014aa}.
While the relatively large uncertainties obtained by
\citet{Kamitsukasa:2014aa} did not permit to detect differences in the
abundances of the elements in these regions, the spectra and best fit
results presented by \citet{Garcia:2012fk} clearly show (at a
7-10~$\sigma$ level) that the presence of the Fe line is much stronger
in the NE and SW regions (Fe/[Fe$_\odot$]=0.09 $\pm0.01$ and 0.12 $\pm0.01$,
respectively) compared to the other regions analysed
(Fe/[Fe$_\odot$]=0.02 $\pm0.01$; the regions used by \citet{Garcia:2012fk}
are reported also in Fig.~\ref{fig:largeMosa}), although no strong
variation between the direction towards (SW) and against (NE) the
initial kick could be seen.

\section{Conclusions}
\label{sec:conclusions}

The overall structures of the PWN and main/counter-jet are confirmed in
the new deep \chan\ observations.  The PWN appears now
more clearly characterised by a bimodal shape, with a wide flow up to
0.7\arcmin\ from the pulsar and a more collimated region extended up to
1.7\arcmin.  We observed a clear softening of the spectrum along the
PWN, and retrieve a nebular magnetic field of $B_{\rm PWN} \gtrsim 30\mu$G.

The main jet still presents overall the same helical pattern seen in the
previous observation,  
although several features at small and large scales cannot be fully explained
in this model. A significant broad emission is detected
around it, distributed asymmetrically with respect to the main jet axis.
Different tentative interpretations of this outflow have been drawn in
Sect.~\ref{sec:discussion}, in particular in the frame of a ballistic
jet, possibly with kink instabilities, and in the frame of diffusion of particles
in the local ISM magnetic field. None of them however 
can reproduce satisfactorily all observed spatial and spectral
characteristics of the main jet, and the full
interpretation of this feature remains open.
A further hint for the presence of multiple flows is even more puzzling.

The lack of \Halpha\ detection (given the presence of strong
surrounding nebulosity) and of detectable proper motion of the pulsar
(given the relatively short timescale over which we could observe the
pulsar, and its large distance to us) are both still fully compatible
with the high linear speed assumed for this pulsar.

In Sect.~\ref{sec:discussion} we also showed that the spectral
properties of MSH~11-61A, derived by 
\cite{Garcia:2012fk},
seem to agree with the general expectation drawn
from the 3D simulations by \citet{Wongwathanarat2013}, and with the
prevailing picture of \j11\ being formed during the same explosion
that produced \SNR.

\begin{acknowledgements} The data analysed in this paper are based on
  \chan\ and VLT observations obtained by our group.

  This research has made use of software provided by the \chan\ X-ray
  Center (CXC) in the application packages CIAO, ChIPS, and Sherpa;
  the Heasarc ftools provided by NASA
  \url{http://heasarc.gsfc.nasa.gov/ftools/} \cite{heasarc}; and the
  ATNF Pulsar Catalogue \citep{ATNFpsrCat}.

  We thank M. Capasso, S. Fotopoulou and C. Baldovin-Saavedra for 
  useful discussions.

\end{acknowledgements}

\bibliographystyle{bibtex/aa} \bibliography{bibliography.bib}

\end{document}